%% file: main.tex
\documentclass{article}

\input{settings.tex}

\input{listing_style}

\title{Abstractions-of-Thought: \\Intermediate Representations for \\ LLM Reasoning in  Hardware Design}

\author{%
  Matthew DeLorenzo \\
  Texas A\&M University\\
  College Station, TX \\
  \texttt{matthewdelorenzo@tamu.edu} \\
  \And
  Kevin Tieu \\
  Texas A\&M University \\
  College Station, TX \\
  \texttt{kevin.tieu@tamu.edu} \\
  \AND
  Prithwish Jana \\
  Georgia Institute of Technology \\
  Atlanta, GA \\
  \texttt{pjana7@gatech.edu} \\
  \And
  Piyush Jha \\
  Georgia Institute of Technology \\
  Atlanta, GA \\
  \texttt{piyush.jha@gatech.edu} \\
  \And
  Dileep Kalathil \\
  Texas A\&M University \\
  College Station, TX \\
  \texttt{dileep.kalathil@tamu.edu} \\
   \And
  Vijay Ganesh \\
  Georgia Institute of Technology \\
  Atlanta, GA \\
  \texttt{vganesh45@gatech.edu} \\
  \And
  Jeyavijayan Rajendran \\
  Texas A\&M University \\
  College Station, TX \\
  \texttt{jv.rajendran@tamu.edu} \\
}

\begin{document}

\maketitle

\begin{abstract}

Large language models (LLMs) have achieved impressive proficiency on logic and programming tasks, often rivaling expert-level performance. However, generating functionally correct hardware description language (HDL) code from natural language specifications remains challenging, primarily in data-scarce domains.

Therefore, we present Abstractions-of-Thought (AoT) — a training-free, inference-only prompting framework to mitigate misinterpretations and reasoning pitfalls of LLMs through a series of task-based abstractions within the prompting procedure, assisting in the transition from high-level to low-level representations of hardware. Furthermore, AoT consists of the following stages: (1) an LLM-based classification of hardware design patterns, (2) a structured intermediate representation (IR) to separate functional decomposition from code syntax, and (3) a line-by-line pseudocode solution enabling a more direct mapping to the final Verilog implementation.  Experimental results on the VerilogEval benchmark depict that AoT demonstrates improvements in functionality when applied to large non-reasoning models (such as \texttt{GPT-4o}), outperforming all baseline techniques (including 1-shot, Chain-of-Thought, and Tree-of-Thought) while significantly reducing the generated tokens by 1.8-5.2$\times$ compared to popular Tree-of-Thought prompting.
\end{abstract}
\section{Introduction}

The rapid development of Large Language Models (LLMs) has driven significant performance improvements in text analysis and generation across various tasks. These advancements are primarily derived from the increases in model size (number of parameters) and the availability of training data and computing power~\citep{minaee2025largelanguagemodelssurvey, zhao2025surveylargelanguagemodels,wei2022emergentabilitieslargelanguage}. In particular, these models have demonstrated strong capabilities in the software industry through automating high-quality code generation, accelerating the software development lifecycle, and enhancing productivity \citep{jiang2024survey}. This has led to the exploration of LLM applications within the chip design process~\citep{liu2023chipnemo, blocklove2023chip, wang2024llmsfuturechipdesign}. Primarily, there is a strong focus on generating integrated circuit (IC) designs in HDLs, such as Verilog, from natural language specifications — an increasingly complex task for designers as semiconductor technology advances. However, LLMs demonstrate relatively limited performance on hardware design benchmarks, as even state-of-the-art (SOTA) LLMs, including ChatGPT~\citep{hurst2024gpt}, DeepSeek~\citep{zhu2024deepseek}, and Llama~\citep{grattafiori2024llama}, can produce mistakes, or ``hallucinations'' that compromise the syntactic correctness or functionality of the intended circuit design \citep{chang2023chipgptfarnaturallanguage,blockloveEvaluatingLLMsHardware2024}. Siemens has attributed part of this challenge to the under-representation of open-source HDL code in the public domain, compared to popular programming languages like Python, Java, and JavaScript \citep{siemens2024llm}. This limitation has motivated further research into optimizing LLMs for hardware generation, seeking to accelerate and automate the chip design workflow.

Digital design of ICs is a complicated and error-prone process, where hardware designers write HDL code at the register-transfer level (RTL) abstraction to implement the precise functionality of circuits according to provided specifications. Therefore, improving the ability of LLMs to not only interpret the natural language specifications of the circuit (e.g., intended logic, Input/Output, and timing components), but also generate compilable and functionally correct RTL code 
has become a significant research focus within hardware development.

As a result, many studies have explored enhancing pre-trained, open-source LLMs through curating Verilog datasets (from online sources and textbooks) to perform fine-tuning and reinforcement learning procedures~\citep{thakur2024verigen, rtlcoder, zhao2024codev, cui2024origen, veriassist, zhang2024mg}. These approaches have demonstrated significant performance improvements over the baseline equivalents, in some cases even challenging SOTA commercial LLMs~\citep{thakur2024verigen, rtlcoder}. Other works mitigate the resources required for training through developing agent-based frameworks around LLMs~\citep{samiNexusLightweightScalable2025}, seeking to improve performance through extended conversational strategies~\citep{blocklove2023chip}, iterative feedback, and interaction with external tools for quality feedback~\citep{thakur2023autochip}. Furthermore, recent works seek to further isolate the framework to the model itself, relying on self-prompting techniques to improve the hardware generation process~\citep{ping2025hdlcore, vijayaraghavan2024chain, sun2025paradigm}. However, these prompting frameworks remain constrained in the use of external tools or datasets, or have limited integration with existing hardware-specific strategies in the LLM generation process.

To this end, we propose \textit{Abstractions-of-Thought (AoT)}, a training-free and agentless prompting framework to improve the quality of generated Verilog through a series of intermediate representations (IR), minimizing the difficulty in translating from high-level descriptions to low-level hardware.
This approach is derived from the fundamental concept of ``abstraction,'' a central methodology in the hardware design process. Hardware engineers utilize varying levels of design representations, such as block diagrams, code representations, or logic gates, to reduce unnecessary complexity in the design. Over the last five decades, this abstraction-driven methodology has been fundamental for the advancement of hardware in many dimensions, including functionality, scalability, design, energy efficiency, high performance, testing, bug-fixing, and security \citep{agarwal2005foundations, sozzo2022pushing}.
The AoT framework consists of a three-stage process, including a (1) classification stage to identify key hardware-design categories, minimizing unnecessary reasoning paths.
Then, (2) the LLM is prompted to represent the circuit solution in an intermediate structured representation, decoupling the design's logic from its implementation in hardware code. (3) Finally, the LLM generates a line-by-line pseudocode abstraction of the solution, leveraging the natural language capabilities of the model to minimize the complexity of the final Verilog generation.

The core contributions of this work are listed below:

\begin{enumerate}[leftmargin=20pt]
    \item \textbf{AoT:} We introduce a novel prompting framework that utilizes multiple abstractions (IRs) to minimize the reasoning difficulty in translating natural language descriptions to domain-specific solutions, such as hardware designs in Verilog.
    
    \item \textbf{Optimized Performance:} A thorough evaluation utilizing the VerilogEval benchmark demonstrates that AoT outperforms alternative prompting strategies (including Chain-of-Thought and Tree-of-Thought) when utilized on large models (\texttt{GPT-4o}) in compilability and functionality.
    \item \textbf{Multi-LLM Approach:} To address the limitations of small models in deriving abstractions, we implement a multi-model AoT strategy, in which a large model performs the abstractions and a small model performs the final Verilog translation. This combination is demonstrated to exceed the performance of either model's individual capabilities across all inference strategies, with up to a 19.7\% improvement in functionality over baseline prompting.
\end{enumerate}

\section{Background}
\label{sec:background}

\subsection{LLMs in Hardware Design}
Due to the inherent limitations of LLMs in hardware design tasks, recent research has focused on domain-specific fine-tuning of open-source pre-trained models~\citep{wang2024llmsfuturechipdesign}. Approaches such as VeriGen mitigate the lack of open-source hardware training data by curating a dataset from publicly available repositories and textbooks, which is used for supervised pre-training~\citep{thakur2024verigen}. Additional works then improved the dataset curation and formatting for training, notably pairing natural language prompts with ideal hardware designs for instruction-tuning alignment~\citep{zhang2024mg, rtlcoder, cui2024origen, pei2024betterv}. Later works further address limitations in training, proposing datasets with multilevel descriptions~\citep{nadimi2024pyranet}, representations of non-textual designs~\citep{liu2024craftrtl}, and code-to-code translation~\citep{cui2024origen}. Furthermore, reinforcement learning has also demonstrated success through numerical feedback based on the quality of the generated hardware design~\citep{veriseek, wang2025insights}. Alternatively, training-free LLM-based frameworks were proposed to mitigate the need for fine-tuning. AutoChip demonstrates that iterative prompting with external simulator feedback effectively refines the final hardware design~\citep{thakur2023autochip}, along with other agent-based approaches for tool-assisted verification ~\citep{veriassist, verilogcoder, sami2024eda}.
Other works improve the prompt strategy, incorporating self-planning into the LLM response for hardware tasks~\citep{lu2024rtllm, vijayaraghavan2024chain} and decomposing large designs~\citep{nakkab2024rome}. Some techniques leverage hardware-specific representations into the prompts, such as ~\citep{sun2025paradigm}, which utilizes intermediate structures for sequential and combinational circuits to assist in Verilog translation. HDLCoRE integrates hardware knowledge through prompt-based self-verification and RAG integration~\citep{ping2025hdlcore}.

Although these works demonstrate improvements, they contain several limitations, including reliance on external tools for feedback, assistance through external databases and retrieval strategies, or minimal integration of hardware design knowledge within the prompt structure. Therefore, we propose a purely inference-based framework that mitigates external dependencies while enabling functionality improvements in LLM-generated Verilog through a series of prompt-driven hardware abstractions, minimizing the transition between natural language and RTL code at each stage.

\subsection{Reasoning in LLMs}
Although LLMs continue to improve, even SOTA models can struggle in extended reasoning tasks, such as multi-step arithmetic or common sense ~\citep{weiChainofThoughtPromptingElicits2023} Therefore, ~\citep{weiChainofThoughtPromptingElicits2023} proposed the Chain-of-Thought (CoT) prompting strategy in which an exemplar (step-by-step thought process for a task) is provided within the prompt context, inducing the model to follow a similar ``reasoning'' structure within its response, demonstrating SOTA performance on reasoning tasks. Furthermore, prompting methods including Tree-of-Thought (ToT)~\citep{yaoTreeThoughtsDeliberate2023} and Graph-of-Thought~\citep{Besta_2024}, were then developed to implement planning capabilities through backtracking and intermediate feedback. Additionally,~\citep{hong2024abstraction} demonstrates that abstraction can elicit improvements in LLM reasoning through addressing complex problems from multiple levels of complexity, gradually refining the solution from an initial high-level representation.
Nevertheless, these inference approaches often incur significant overhead in the number of generated tokens due to verbose or inefficient intermediate reasoning paths. To minimize these paths, works including Chain-of-Draft~\citep{xu2025chain} and Sketch-of-Thought (SoT) \citep{aytes2025sketch} demonstrate that reasoning tokens can be condensed with minimal degradation to performance on reasoning tasks, including mathematics and logic.

However, these initial reasoning approaches offer limited optimizations regarding reasoning tasks similar to hardware code design, which presents unique challenges. This includes an absence of foundational training in hardware design principles and processes. Moreover, the verbosity of reasoning motivates the need for more concise and structured reasoning representations, especially when applied to tasks like large-scale hardware code. Given the initial success of abstraction in LLM reasoning~\citep{hong2024abstraction}, our AoT framework addresses hardware domain challenges by leveraging alternative representations frequently utilized in circuit design, commonly referred to as ``hardware abstractions,'' effectively integrating a reasoning structure for Verilog code generation.

\section{Framework} \label{sec:framework}
\begin{figure*}[t]
    \centering
    \includegraphics[width=\textwidth,trim={0 0cm 0 0cm},clip]{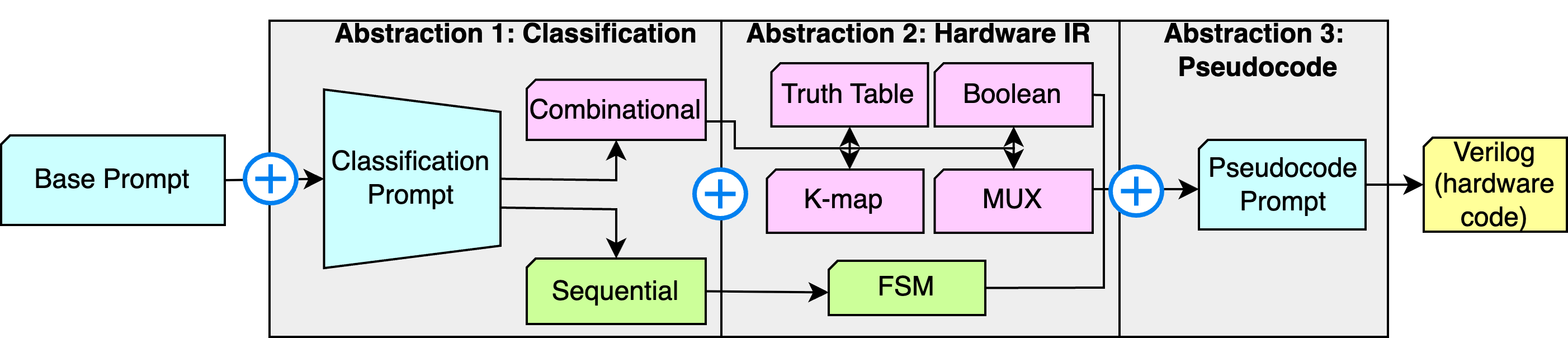}
    \caption{AoT Framework — abstractions for generating hardware designs.}
    \vspace{-1.7em}
\label{fig:final_framework_figure}
\end{figure*}

As the semiconductor industry continues to push the limits of transistor technology, computer chips have become faster, more power-efficient, and increasingly complex \citep{burkacky2022mckinsey}. These advancements, coupled with the industry pressures to continuously deliver improved performance, induce significant security risks as undetected defects or alterations throughout chip design continue to grow. As a result, chip manufacturers have begun investing heavily in validation and verification to ensure design integrity and quality \citep{wang2024llmsfuturechipdesign}. In managing growing design complexity and workloads, hardware designers must approach the design process through varying levels of detail such that specific tasks can be effectively addressed while minimizing unnecessary complexity. This is accomplished through hardware abstraction, ranging from high-level circuit specifications to transistor-level circuit placement \citep{agarwal2005foundations, sozzo2022pushing}. This concept of abstraction is widely utilized across the chip design industry to optimize for various chip design objectives, including low-power optimization, functionality testing, security, and validation~\citep{yoo2003introduction, sozzo2022pushing, chippa2014scalable}. 

Within the design stage, these abstractions serve various purposes. For instance, for rapid, large-scale design prototyping, designers utilize architectural-level abstractions to define blocks and interfaces representing large computing components (i.e., processors, memory, and interconnects). Then, the behavioral abstraction defines the functionality of components with high-level synthesis tools, enabling system-level verification and architecture trade-offs. The hardware-code abstraction (e.g., Verilog) allows the circuit's logic to be directly defined on each clock edge, allowing designers to perform formal verification and catch bugs before moving to lower-level implementations~\citep{herklotz2021formal}. These abstractions simplify complex tasks through an iterative, high-level to low-level approach~\citep{sozzo2022pushing}.

Akin to designers, LLMs benefit from well-defined, manageable tasks step-by-step. We therefore propose the AoT framework to assist in the inference process. Through abstracting the original hardware description prompt into multiple intermediate formats, we aim to concentrate the natural language processing and problem-solving capabilities of the LLM on transitional tasks that assist in determining the final Verilog translation, thereby improving the LLMs intermediate reasoning procedure. As illustrated in Figure~\ref{fig:final_framework_figure}, the AoT approach consists of a series of abstractions (including classification, problem-specific IR, and pseudocode) that effectively address limitations of LLM reasoning capabilities in hardware design tasks, enabling a novel thinking paradigm for high-quality Verilog generation. The implementation of each abstraction layer is further defined in detail below.

\subsection{High-Level Abstraction: Module Classification}\label{sec:abstraction1}
To implement a hardware design, designers often begin by first classifying the intended circuit functionality by common design patterns or structures, enabling existing strategies to assist in the implementation process~\citep{ping2025hdlcore, sun2025paradigm}. Furthermore, two primary classifications of fundamental circuit design include \textit{combinational} and \textit{sequential} circuits. These categories provide insight into the circuit's necessary components based on their functional dependencies. This includes that combinational circuits depend solely on the current value of their inputs, with no memory of prior states (e.g., an adder module). However, the functionality of a sequential circuit is not only dependent on its current input values, but also on its current state stored in a memory component (e.g., a counter module). Therefore, the first-level abstraction of AoT utilized the LLM to perform a two-stage classification procedure to determine the optimal structure of the circuit design described in the prompt, thereby minimizing incorrect reasoning paths and providing insight into later abstractions.

To implement the first stage of this abstraction, it is assumed that we have a set of base prompts that briefly describe a Verilog design in natural language. In this work, we utilize the VerilogEval prompts, further detailed in Section~\ref{sec:benchmark}. To then classify the structure of each hardware design, a template prompt is constructed that directs the LLM to analyze the given design and determine if the functionality should be combinational or sequential, defined as $C_1$. To assist in this process, this prompt incorporates additional hardware knowledge necessary for this task through brief definitions of each category, along with circuit components indicative of the classification (i.e., clock signals, memory components, and operation types). 
The LLM is then prompted to respond with a single word final answer, ``combinational'' or ``sequential,'' resulting in a set of initial classifications.

Upon the delineation above, a second classification prompt ($C_2$), is introduced. This component seeks to further refine the most optimal intermediate design strategy based upon the base design prompt and its prior classification. For instance, designers often utilize an intermediate structure to effectively represent the circuit's functionality before directly implementing it in hardware code. In the case of sequential circuits, this structure is directly defined as a \textit{finite-state-machine diagram} (FSM), representing all 
possible states and transitions. However, combinational circuits have a larger variety of useful representations utilized by designers for defining 
\begin{wrapfigure}{r}{0.48\columnwidth}
  \vspace{-1.5em}
  \begin{tcolorbox}[snippetbox]
    \begin{lstlisting}[style=SnippetStyle]
Classify the combinational design.
- truth table
- Boolean expression
- karnaugh-map
- MUX
- other

Advantages/disadvantages:
- truth_table: Effective in enumerating all possible outputs with minimal inputs...

- boolean_expression: Effective for algorithmic or pattern-based logic given many inputs... 

- K-map: Effective for minimizing logic expressions to minimize complexity...

- mux_mapping: Effective when selecting among several data inputs using select signals...
    \end{lstlisting}
  \end{tcolorbox}
  \vspace{-1em}
  \caption{Classification prompt ($C_2$). }
  \vspace{-2em}
  \label{fig:classification_prompt}
\end{wrapfigure}
the logic of a circuit, dependent on the design complexity and objective. Therefore, if 
the circuit is combinational, the $C_2$ prompt classifies the circuit according to a representative set of typical combinational structures, including: \textit{truth tables},\textit{ boolean expressions}, \textit{Karnaugh-Maps}, and \textit{multiplexers (MUX)}, each of which is further detailed in Section~\ref{abstraction2}. This prompt, similar to $C_1$, assists the LLM in the classification process by incorporating hardware-design knowledge, consisting of the advantages and disadvantages of each representation, as shown in Figure~\ref{fig:classification_prompt}. An additional condition, ``other,'' is included if the LLM determines that the design task is not represented, in which case the abstraction skips directly to final abstraction (Section~\ref{sec:pseudocode_abstract}).

With this high-level classification obtained, the determined classification and structure category can then be appended to the baseline prompt to direct the LLM towards the most effective approach in determining the Verilog implementation. More importantly, the determined classification is then utilized within the next abstraction layer to construct a more detailed representation of the design.

\subsection{Mid-Level Abstraction: Problem-Specific Intermediate Representation (IR)}\label{abstraction2} 

Given that the Verilog module has now been classified into its optimal intermediate structure from the first abstraction stage (Section~\ref{sec:abstraction1}), this abstraction seeks to define a lower-level representation of the circuit's logic by constructing a text-based intermediate representation (IR) of the design based upon the previously defined structure. 
These structures enable the intended functionality of the described circuit to be fully represented in a condensed format as opposed to hardware code, similar to short-hand notations utilized by domain-experts to enable effective reasoning~\citep{aytes2025sketch}. 
Notably, this ability to represent circuits through abstractions enables us to separate the task of solving the logical implementation of the circuit from the task of Verilog programming. Therefore, the natural language and logical capabilities of LLMs can be further leveraged without the dependence on a thorough understanding of hardware code syntax for effective representation.
\begin{wrapfigure}{r}{0.48\columnwidth}
  \vspace{0em}
  \begin{tcolorbox}[snippetbox]
    \begin{lstlisting}[style=SnippetStyle]
FSM-JSON
{
  "states": ["S1", ..., "S10"],
  "transitions": [
    {"from": "S1", "to": "S1", "cond": "reset"},
    {"from": "S1", "to": "S2", "cond": "!reset"},
    ...
    {"from": "S10", "to": "S1", "cond": "reset"},
    {"from": "S10", "to": "S1", "cond": "!reset"}
  ],
  "outputs": [
    {"state": "S1", "signal": "q", "value": 1},
    ...
    {"state": "S10", "signal": "q", "value": 10}]
}
    \end{lstlisting}
  \end{tcolorbox}
  \vspace{-1em}
  \caption{Counter module IR abstraction.}
  \vspace{-2em}
  \label{fig:ir_counter}
\end{wrapfigure}

In implementing this abstraction, we seek to translate each of the five structures into a text-based format easily interpretable by the LLM. Therefore, motivated by the paradigm approach~\citep{sun2025paradigm}, we provide various JSON formats that represent each of our structures to ensure an organized and concise representation of the information. This is accomplished through five template prompts that provide the associated JSON format, directing the LLM to implement the circuit functionality accordingly. The definitions of each structure included in the templates are listed below, along with the information captured in the JSON string.

\textbf{FSM.} utilized for all sequential circuits to delineate all possible states and transition conditions for the circuit, and the associated output functionality. The JSON list includes: the states, transitions, and outputs (see Figure~\ref{fig:ir_counter}).
\textbf{FSM.} utilized for all sequential circuits to delineate all possible states and transition conditions for the circuit, and the associated output functionality. The JSON list includes: the states, transitions, and outputs (see Figure~\ref{fig:ir_counter}).

\textbf{Truth Tables.} If the circuit contains a small number of input parameters (\(\leq 4\)), the LLM is prompted to define all possible logical combinations in a tabular format. The JSON list includes: each input combination and the associated output.

\textbf{Boolean Equations.} If the circuit contains a larger number of input parameters, where truth table representations are impractical, the LLM is directed to express the final logic in terms of Boolean equations. The JSON list includes: the inputs variables, output variables, and logic expressions.

\textbf{Karnaugh Map.} In the case that the circuit design problem requires logic minimization, the Karnaugh Map (K-map) structure is utilized to simplify the logic expressions. The JSON list includes: column variables, row variables, and K-map values.

\textbf{MUX.} If the problem requires selecting data based on input parameters and selection conditions, the data-routing hardware component, MUX, is leveraged. The JSON list include: input parameters, select variable, and associated output.

Upon the generation of the JSON representation, the output IR is then combined with the base prompt for either direct evaluation, or for utilization in the final abstraction component.

\subsection{Low-Level Abstraction: Line-by-Line Pseudocode}\label{sec:pseudocode_abstract}
After the classification and IR stages, the final abstraction seeks to provide a low-level, line-by-line pseudocode representation of the associated Verilog module. This step enables the natural language processing capabilities of LLMs to be leveraged through defining the implementation in greater detail than prior abstractions, while still mitigating the constraints of hardware-specific Verilog syntax. Additionally, this abstraction simplifies the final translation process to Verilog to a series of smaller operations, thereby minimizing the complexity of the final RTL code generation.

\begin{figure}[H]
   \vspace{-0.5em}
  \begin{tcolorbox}[snippetboxfull]
    \begin{lstlisting}[style=SnippetStyle]
 Module declaration: Define module "top_module" with port "clk" (1-bit input), "reset" (1-bit input, synchronous active-high), and "q" (4-bit output register)
 Create always block triggered on the rising edge of "clk" for sequential logic
 Within the always block, check if "reset" is asserted (i.e., high)
     If true, assign q <- 1 to reset the counter to the initial state (count 1)
 Otherwise (reset is not asserted), check if q equals 10 (the maximum count)
     If true, assign q <- 1 to wrap the count back to 1
     Else, assign q <- q + 1 to increment the counter by one
 End the always block and conclude the module definition
    \end{lstlisting}
  \end{tcolorbox}
   \vspace{-1em}
  \caption{Counter module pseudocode abstraction.}
  \vspace{-1em}
  \label{fig:inline-snippet}
\end{figure}

This component is implemented similarly to the prior abstractions, in which the template prompt is first defined. Then, for a given design, the standard prompt and any of its prior abstraction results (i.e., classification and IR) are passed into the template prompt. This template prompt instructs the LLM to analyze the design and abstractions, and then provide a line-by-line pseudocode of the associated Verilog module, without generating any actual Verilog code. Upon the extraction of the isolated pseudocode (Figure~\ref{fig:inline-snippet}), it is then appended to the existing prompt and prior abstractions, resulting in the final AoT prompt. This is then passed to the LLM to generate the final Verilog module translation.

\section{Experiments}\label{sec:experiments}
We perform a comprehensive evaluation of the Abstractions-of-Thought framework above to address the following three key research questions (\textbf{RQs}):

\begin{enumerate}[label=\textbf{RQ\arabic*.}, leftmargin=30pt]
\item How does AoT compare to alternative prompting strategies for hardware design across models of varying sizes and capabilities?

\item Given that the AoT framework consists of multiple abstractions, can a multi-model approach elicit additional performance improvements?

\item Which abstractions within the AoT framework are most effective in improving functionality?

\end{enumerate}

\subsection{Evaluation Benchmark: VerilogEval}\label{sec:benchmark}
For a comprehensive metric that quantifies LLM performance in hardware design, we utilize the VerilogEval-Human v1.0.0 benchmark \citep{liu2023verilogeval}. This evaluation contains of 156 prompts, each consisting of a natural language description of the functionality of a Verilog module, followed by the module's instantiation (name and I/O parameters). From these prompts, the LLM is then expected to complete the implementation of the design in Verilog. VerilogEval then evaluates the generated modules by simulation (Icarus Verilog) to determine if they are compilable. Additionally, each module is validated against its associated testbench vector to determine functional correctness. For these metrics, the results are defined by pass@k, an unbiased estimator for the likelihood that at least one of the $k$ selections from $n$ samples is correct, where $c$ is the total number correct completions of $n$ samples, defined as: pass@k = $\mathbb{E}[(1 - C(n-c, k)), C(n, k)]$ \citep{chenEvaluatingLargeLanguage2021}.

\subsection{Experimental Configuration and Procedure}\label{sec:experiment_params}
Across all LLM experiments defined below, we maintain consistent hyperparameters during the inference process. The temperature is set to a predefined value of 0.6 out of 2.0 to ensure variability in the generated modules, along with a top-k of 0.99. All other hyperparameters, such as maximum output tokens, are left to their default values. To perform inferencing across a variety of model types, we utilize the associated API's for closed-source models (i.e., OpenAI). For open-source models, we leverage the Huggingface Endpoint platform for model deployment, enabling sufficient GPU resources (the largest being an NVIDIA L40S) to be allocated for our varying inference tasks.

\textbf{RQ1}. The AoT framework is evaluated on all VerilogEval prompts ($n=5$) across a representative set of SOTA LLMs, varying in size, capability, and accessibility. Through these evaluations, we can then identify where the AoT framework is most effective relative to alternative inference frameworks, including 1-shot, CoT, SoT, and ToT. These models consist of:
(1) \texttt{DS-Coder-V2-Lite-Instruct}: DeepSeek's 16B parameter open-source model fine-tuned for programming tasks~\citep{zhu2024deepseek}.
(2) \texttt{Llama-3.1-8B-Instruct}: an instruction-tuned open-source model for general tasks~\citep{grattafiori2024llama}.
(3) \texttt{GPT-4o}: OpenAI's flagship model, versatile in many modalities.
(4) \texttt{GPT-4o-mini}: a smaller \texttt{GPT-4o} model optimized for speed and cost~\citep{hurst2024gpt}.

\textbf{RQ2.} To evaluate the multi-model implementation of AoT, we perform the same evaluation process as above, but employ different models for the abstraction and the final Verilog translation stages. Specifically, a large foundational model \texttt{GPT-4o-mini} is used to generate the intermediate abstractions (pseudocode and structural IRs), while a small coding model is used for the final translation to Verilog. For fair comparison, the same multi-model approach is performed for the other strategies when applicable (including CoT and ToT), in which the reasoning prompts can be separated from the final generation. Other single-prompt methods (baseline, one-shot, and SoT) cannot be separated into a multi-model strategy, therefore their performance is based on the single model utilized for Verilog generation, denoted in Table~\ref{tab:GroupedVerilogEval_pass1}. Through this evaluation, we can evaluate how the quality of the intermediate abstractions impacts the final hardware design translation.

\textbf{RQ3.} In defining the contributions of each AoT component to final Verilog generation capabilities, an ablation study is conducted across each abstraction layer combination of AoT. This includes the following configurations: Base prompt (VerilogEval), only pseudocode, only structural IR, Base + Structural IR, Base + Pseudocode, and Base + IR + Pseudocode. Each of these abstraction combinations is generated by \texttt{GPT-4o-mini}, coupled with both the \texttt{DS-Coder-V2-Lite-Instruct} and \texttt{Llama-3.1-8B-Instruct} model for the final Verilog generation to examine the effectiveness of each abstraction in improving functionality.

\section{Results}\label{sec:results}

{
\scriptsize
\setlength{\tabcolsep}{2pt}
\begin{table*}[h!]
  \caption{Evaluation of LLMs on VerilogEval across Prompt Strategies (\textit{Asterisk (*) denotes single-model results for comparison due to incompatibility for multi-model inference strategies})}
  \label{tab:GroupedVerilogEval_pass1}
  \centering
  \resizebox{0.8\textwidth}{!}{%
    \begin{tabular}{cc*{5}{c}c}
      \hline
      \multirow{2}{*}{Model} & \multirow{2}{*}{Evaluation}
        & \multicolumn{6}{c}{Accuracy (\%)} \\
      \cline{3-8}
      & 
        & Baseline & 1‑shot & CoT   & SoT   & ToT   & AoT (ours)\\
      \hline
      
        \multirow{2}{*}{GPT‑4o}
          & Compilation
            & 72.3    & 71.7  & 78.3    & 75.4  & 77.4    & \cellcolor{blue!20}\textbf{80.9} \\
          & Functionality
            & 57.8    & 56.2  & 59.0    & 56.3  & 60.1  & \cellcolor{blue!20}\textbf{60.4 }\\
      \hline

      \multirow{2}{*}{GPT‑4o‑mini}
        & Compilation
          & 67.1    & 69.5  & 69.9    & 12.6  & 62.8    & \cellcolor{blue!20}\textbf{74.9} \\
        & Functionality
          & \cellcolor{blue!20} \textbf{48.3}  & 48.1  & 46.2    & 6.9   & 40.8    & 47.7 \\
      \hline        

      \multirow{2}{*}{DS-Coder-V2-Lite-Instruct}
        & Compilation
          & 79.9    & 72.2  & \cellcolor{blue!20}\textbf{80.6}  & 70.9  & 77.1    & 76.3 \\
        & Functionality
          & 46.9    & 43.8  & \cellcolor{blue!20}\textbf{49.7}  & 44.2  & 45.3    & 40.4 \\
      \hline

      \multirow{2}{*}{Llama‑3.1‑8B-Instruct}
        & Compilation
          & 41.0    & 31.9  & \cellcolor{blue!20}\textbf{52.7}  & 23.8  & 39.1    & 47.9 \\
        & Functionality
          & 16.2    & 13.1  & \cellcolor{blue!20}\textbf{21.9}  & 7.1   & 9.62    & 13.7 \\
      \hline

      \multirow{2}{*}{4o‑mini \& DS-Coder-V2-Lite-Instruct}
        & Compilation
          & 79.9\textsuperscript{*}    & 72.2\textsuperscript{*}   & 79.3           & 70.9\textsuperscript{*}   & 78.3    & \cellcolor{blue!20}\textbf{80.1} \\
        & Functionality
          & 46.9\textsuperscript{*}    & 43.3\textsuperscript{*}   & 50.4           & 44.2\textsuperscript{*}   & 47.3    & \cellcolor{blue!20}\textbf{51.5} \\
      \hline

      \multirow{2}{*}{4o‑mini \& Llama‑3.1‑8B-Instruct}
        & Compilation
          & 41.0\textsuperscript{*}    & 31.9\textsuperscript{*}   & 52.8           & 23.8\textsuperscript{*}  & 51.9   & \cellcolor{blue!20}\textbf{68.5} \\
        & Functionality
          & 16.2\textsuperscript{*}        & 13.1\textsuperscript{*}      &       25.4         & 7.1\textsuperscript{*}     & 14.1        &   \cellcolor{blue!20}\textbf{35.9}   \\
      \hline
      
    \end{tabular}
  }
\end{table*}
}

\begin{table*}[h!]
  \caption{Average Generated Tokens across Strategies on VerilogEval-Human (\textit{Asterisk (*) denotes single-model results for comparison due to incompatibility for multi-model inference strategies})}
  \label{tab:TokenUsage}
  \centering
  \resizebox{0.9\textwidth}{!}{%
    \begin{tabular}{c*{6}{c}c}
      \hline
      Model
        & Baseline & 1‑shot & CoT   & SoT   & ToT   & AoT (ours)  & AoT (per abst.) \\
      \hline
      GPT‑4o
        & {501} & {376} & {491} & {200}  & {2213}    & {886}  & {295} \\
      GPT‑4o‑mini
        & {623} & {502} & {609} & {200}  & {2671}    & {1018} & {339} \\
      DS‑Coder‑V2‑Lite‑Instruct
        & {601} & {420} & {497} & {349}  & {2455}    & {476}  & {159} \\
      Llama‑3.1‑8B
        & {535} & {503} & {546} & {482}  & {3244}    & {1762} & {587} \\
      4o‑mini \& DS-Coder-V2-Lite-Instruct
        & 601*   & 420*  & 398   & 349*    & 2426      & 1046   & 349   \\
      4o‑mini \& Llama‑3.1‑8B-Instruct
        & 535*   & 503*   & 457   & 482*    & {2383}    &   1131     &   377   \\
      \hline
    \end{tabular}%
  }
\end{table*}

\begin{table*}[h!]
  \caption{Ablation Study on AoT Components on VerilogEval-Human (\textit{Asterisk (*) denotes single-model results for comparison due to incompatibility for multi-model inference strategies})}
  \label{tab:ablation}
  \centering
  \resizebox{\textwidth}{!}{%
    \begin{tabular}{cc*{6}{c}}
      \hline
      \multirow{2}{*}{Model} & \multirow{2}{*}{Evaluation}
        & \multicolumn{6}{c}{Accuracy (\%)} \\
      \cline{3-8}
      & 
        & Base & Pseudo & Base + Pseudo & IR
        & Base + IR & Base + IR + Pseudo \\
      \hline
      
      \multirow{2}{*}{4o‑mini \& DS-Coder-V2-Lite-Instruct}
      & Compilation
        & 79.9\textsuperscript{*} & 69.7 & 78.1 & 57.8 & 78.3 & \cellcolor{blue!20}\textbf{80.1} \\
      & Functionality
        & 46.9\textsuperscript{*} & 47.1 & 50.6 & 21.4 & 45.9 & \cellcolor{blue!20}\textbf{51.5} \\
      \hline

      \multirow{2}{*}{4o‑mini \& Llama‑3.1‑8B-Instruct}
      & Compilation
        & 41.0\textsuperscript{*} & 67.1 & 68.1 & 45.6 & 44.2 & \cellcolor{blue!20}\textbf{68.5} \\
      & Functionality
        & 16.2\textsuperscript{*} & \cellcolor{blue!20}\textbf{35.9} & 32.4 & 16.4 & 18.7 & 33.2\\
      \hline

    \end{tabular}%
  }
\end{table*}

\subsection{AoT Performance across Models (RQ1)}
After evaluating the optimal configuration of the AoT strategy against alternative approaches, the results across the VerilogEval benchmark (depicted in Table~\ref{tab:GroupedVerilogEval_pass1}) demonstrate the following findings. Regarding \textbf{RQ1}, we first observe that the AoT approach outperforms all other strategies on 1 out of 4 base models (\texttt{GPT-4o}) in the functionality rate, providing a 2.6\% increase over baseline prompts and a 0.3\% increase over ToT, the next best approach. Conversely, for the remaining models, AoT is outperformed by at least one alternative prompting strategy, with particular limitations when applied to smaller models (\texttt{DS-Coder-V2-Lite-Instruct} and \texttt{Llama‑3.1‑8B-Instruct}), in which AoT does not outperform baseline prompting. Therefore, our observations indicate that AoT demonstrates the most efficacy in larger models that are flexible across varying NLP tasks, enabling high-quality abstractions to assist in the Verilog generation process. Additionally, we observe the generated tokens of each strategy in Table~\ref{tab:TokenUsage}, indicating that AoT incurs an overhead over most strategies (detailed in Section~\ref{sec:limitations}). However, we do find that AoT reduces the generated tokens compared to ToT by a factor of 1.8-5.2$\times$, demonstrating an optimal tradeoff in hardware design tasks. 

\subsection{AoT Performance with Multi-Model Approach (RQ2)}\label{sec:multimodal}
In evaluating the efficacy of AoT in the multi-model approach \textbf{(RQ2)}, the results in Table~\ref{tab:GroupedVerilogEval_pass1} demonstrate that the Verilog generation quality of both smaller models is significantly improved with AoT when the abstractions are generated from larger LLMs. With \texttt{GPT-4o-mini} generated abstractions, \texttt{DS-Coder-V2-Lite-Instruct} achieves an 11.1\% improvement in its functional accuracy (51.5\%) over its single model AoT equivalent, while also exceeding all alternative prompting approaches. The \texttt{Llama‑3.1‑8B-Instruct} model further supports this trend, attaining a 22.2\% improvement over its single-model AoT, and exceeds its prior best single-model strategy (CoT) by 14\%. Additionally, the performance of \texttt{DS-Coder-V2-Lite-Instruct} with \texttt{GPT-4o-mini} generated abstractions is improved over not only \texttt{DS-Coder-V2-Lite-Instruct}, but also outperforms the \texttt{GPT-4o-mini} model itself. These results not only emphasize the importance of abstraction quality when applying the AoT framework to the final result, but also demonstrate that AoT with a multi-model setup can exceed the individual performance of either model's capabilities.

\subsection{Ablation Study on Abstractions (RQ3)}\label{sec:ablation}
To address \textbf{RQ3}, we conduct an ablation study by applying each component of the AoT framework to the two multi-model configurations: \texttt{GPT-4o-mini} with \texttt{DS-Coder-V2-Lite-Instruct} and \texttt{Llama‑3.1‑8B-Instruct}. The results, shown in Table~\ref{tab:ablation}, reveal that using IR abstractions alone (without the base prompt) is ineffective for either model, notably causing a 25.5\% drop in functionality for the \texttt{DS-Coder-V2-Lite-Instruct} configuration. In contrast, combining IR with the base prompt improves the performance in both setups, surpassing the baseline functionality by 2.5\% for the Llama configuration.
Regarding the pseudocode abstraction, using it without the base prompt is beneficial in both configurations, doubling the baseline functionality of the \texttt{Llama‑3.1‑8B-Instruct} configuration (from 16.2\% to 35.9\%). Furthermore, combining the pseudocode with the base prompt further enhances the compilability of both models. The best results are achieved by utilizing the full AoT framework on the \texttt{DS-Coder-V2-Lite-Instruct} configuration, resulting in a 4.6\% gain in functionality over the baseline and the overall most effective configuration. Moreover, these results demonstrate that both abstractions perform best with the base prompt, while pseudocode proves to be a more effective abstraction than IR. Furthermore, utilizing the full AoT framework demonstrates the highest functionality accuracy, reinforcing the efficacy of combining multiple abstractions.

\section{Limitations}\label{sec:limitations}
Although the AoT framework improves performance for large models (\texttt{GPT-4o} and \texttt{GPT-4o-mini}), its effectiveness diminishes when applied to the smaller models (\texttt{DS-Coder-V2-Lite-Instruct} and \texttt{Llama‑3.1‑8B-Instruct}). This suggests that the quality of the intermediate abstractions (pseudocode and IR) generated by the LLM plays a critical role in AoT's efficacy. To validate this, we conduct an ablation study (Section~\ref{sec:ablation}), showing that the abstractions generated by a larger model (\texttt{GPT-4o-mini}) enable a smaller model to outperform its baseline by up to 2$\times$.
Therefore, this observation can be attributed in part to the inherent limitations of the base models, as the quality of each abstraction component is directly dependent on the NLP capabilities of the model. Furthermore, the full AoT strategy introduces overhead in token usage (Table~\ref{tab:TokenUsage}), producing 0.8-3.3$\times$ the number of tokens from baseline prompting. This is primarily due to the multi-stage prompt structure of AoT.

While VerilogEval serves as the de facto benchmark for this line of research, it has notable limitations. It targets single-module designs and lacks evaluations on large-scale hardware designs (e.g., microprocessors, SoCs, or accelerators), leaving the scalability of AoT on complex designs untested. However, AoT is well-suited for further extension, as its modular structure supports integration of additional abstractions for complex design objectives.

\section{Conclusion}
We introduce Abstractions-of-Thought (AoT), a novel LLM prompting strategy that outperforms existing techniques, such as CoT, SoT, and ToT, in generating hardware designs. The superiority of AoT stems from its structured, three-stage abstraction mechanism that progressively refines high-level descriptions into detailed low-level solutions—an approach inherently aligned with the hierarchical nature of hardware design. Unlike conventional one-prompt strategies, AoT's multi-stage abstraction process optimally aligns with large models like \texttt{GPT-4o} and \texttt{GPT-4o-mini}, where its normalized performance overhead is minimal. As a result, AoT provides superior functionality and efficiency in complex reasoning tasks, while also reducing the token overhead by up to 5.2$\times$ over ToT. The AoT methodology is not limited to hardware and has the potential to enhance reasoning in other engineering domains that employ multi-layered abstractions, such as software engineering and cyber-physical system design. Thus, AoT establishes a compelling framework for leveraging LLMs in complex, abstraction-heavy design tasks.

\section{Acknowledgment}
The authors acknowledge the support from the Purdue Center for Secure Microelectronics Ecosystem — CSME\#210205.
\bibliographystyle{abbrvnat}
\bibliography{main.bib}

\newpage
\appendix

\section{Technical Appendices and Supplementary Material}
Within the Appendix below, additional material is included regarding the AoT framework and associated experiments. First, the AoT framework is clearly defined in a mathematical description for clarity in the structure of the AoT process. Then, the template prompts and intermediate abstractions defined in AoT are demonstrated through example listings. Additionally, an extended set of experiments is then included (including pass@5 metrics and additional model evaluations), along with the statistical significance of the primary pass@1 results (including compilability, functionality, and tokens).
\subsection{Framework — Mathematic Notation}\label{sec:math_notation}

We first provide a concise mathematical description of the entire Abstraction‑of‑Thought (AoT) prompting procedure. Our goal is to make clear:
\begin{itemize}
  \item the domain and codomain of each template prompt,
  \item the intermediate variables (i.e., abstractions) that it produces, and
  \item how they compose into the final AoT prompt structure for hardware design.
\end{itemize}

\textbf{{Set of standard prompts}}

Let
\[
  S = \{\,s_1,\,s_2,\,\dots,\,s_{156}\}
\]
denote the collection of our 156 original hardware‑design prompts from the VerilogEval-Human (v1.0) evaluation set.

\textbf{{Output spaces}}

We introduce the following target sets:
\begin{itemize}
  \item $C_1$: first‑level classifications (e.g.\ “combinational” vs.\ “sequential”),
  \item $C_2$: second‑level, more specific classifications (e.g.\ “truth table,” “K‑map,” etc.),
  \item $R$: the space of JSON-based intermediate representations (IR),
  \item $P$: the space of pseudocode abstractions, and
  \item $A$: the space of final Verilog‑solution outputs.
\end{itemize}

\textbf{{Template functions}}

We model each prompt template as a mathematical function:
\[
\begin{aligned}
  f_{\mathrm{cls1}} &: S \;\longrightarrow\; C_1, 
    &&\text{(first‑level classification)}\\
  f_{\mathrm{cls2}} &: S \times C_1 \;\longrightarrow\; C_2, 
    &&\text{(refined structural classification)}\\
  f_{\mathrm{IR}}   &: S \times C_1 \times C_2 \;\longrightarrow\; R, 
    &&\text{(structured intermediate representation)}\\
  f_{\mathrm{ps}}   &: S \times C_1 \times C_2 \times R \;\longrightarrow\; P, 
    &&\text{(line‑by‑line pseudocode)}\\
  f_{\mathrm{final}}&: S \times C_1 \times C_2 \times R \times P 
                      \;\longrightarrow\; A,
    &&\text{(final Verilog solution).}
\end{aligned}
\]

\textbf{{Per‑prompt pipeline}}

For each \(i = 1,\dots,156\), we apply these in sequence:
\[
\begin{aligned}
  c_i^{(1)} &= f_{\mathrm{cls1}}\bigl(s_i\bigr),
    &&\text{(is \(s_i\) combinational or sequential?)}\\
  c_i^{(2)} &= f_{\mathrm{cls2}}\bigl(s_i,\;c_i^{(1)}\bigr),
    &&\text{(which specific structure fits best?)}\\
  r_i        &= f_{\mathrm{IR}}\bigl(s_i,\;c_i^{(1)},\;c_i^{(2)}\bigr),
    &&\text{(build the JSON‑style IR)}\\
  p_i        &= f_{\mathrm{ps}}\bigl(s_i,\;c_i^{(1)},\;c_i^{(2)},\;r_i\bigr),
    &&\text{(generate line‑by‑line pseudocode)}\\
  a_i        &= f_{\mathrm{final}}\bigl(s_i,\;c_i^{(1)},\;c_i^{(2)},\;r_i,\;p_i\bigr),
    &&\text{(produce the final Verilog code).}
\end{aligned}
\]

\textbf{{Compact composition}}

All five steps can be seen as one composite mapping:
\[
  F \;=\;
    f_{\mathrm{final}}
    \;\circ\;
    \bigl(\mathrm{id},\,f_{\mathrm{cls1}},\,f_{\mathrm{cls2}},\,f_{\mathrm{IR}},\,f_{\mathrm{ps}}\bigr),
\]
so that simply
\[
  a_i = F\bigl(s_i\bigr)
\quad\text{for each }i.
\]

\subsection{Template Prompts}
Included below is additional information regarding each template prompt structure utilized within the AoT framework. In this section, an example of each prompt template is depicted along with its utilization in the AoT process, with all notable components in each example highlighted in detail. These descriptions will abide by the notations in Section~\ref{sec:math_notation} for clarity.
\begin{lstlisting}[language=Python, label = {listing:listing1}, caption={Baseline VerilogEval prompt format (ex: full adder module).},style=prettyverilog,aboveskip=10pt,firstnumber=1,linewidth=\linewidth]
//Create a full adder. 
//A full adder adds three bits (including carry-in) and produces a sum and carry-out.

module top_module ( 
    input a, b, cin,
    output cout, sum rightbracket;
\end{lstlisting}
Above in Listing~\ref{listing:listing1} is a standard prompt format utilized by VerilogEval-Human (v1.0). As demonstrated with an example of a full adder module, the prompt consists of two primary components — a high-level natural-language description, and the module instantiation (i.e., the module name and I/O parameters). The LLM is then expected to complete the module above with the correct functionality. This prompt structure is a standard procedure in VerilogEval evaluations, and is representative of the baseline approach in LLM-assisted hardware design. This prompt is then utilized within the AoT framework to derive additional information through abstraction for improved Verilog generation.

\begin{lstlisting}[language=Python, label = {listing:listing2}, caption={Classification prompt ($f_{\mathrm{cls1}}$) defining if the design is sequential or combinational.},style=prettyverilog,aboveskip=10pt,firstnumber=1,linewidth=\linewidth]
You are a professional hardware engineer. 
Please read the Verilog description and instantiation below and determine whether its logic is purely combinational or if it contains sequential elements (i.e., flip-flops, registers, clocked always blocks).

------Beginning of High level description of Verilog------

{description}

------End of High level description of Verilog------

Classify the above module as either "combinational" or "sequential". 
Respond with exactly one word in a code block, either:
```combinational```
```sequential```
Please do not respond with any other text in your response.
"""
\end{lstlisting}
Within Listing~\ref{listing:listing2} is the first stage of AoT — the first classification prompt. As shown in the structure above, after an initial instruction defining the objective to the LLM (to identify if the module is combinational or sequential), the prompt from Listing~\ref{listing:listing1} is then included in the {description} section. To ensure a consistent output format, the template defines the LLM to respond in a single final-word format. In our observations, we found that restricting the LLM's maximum token generation to a single word length harmed the accuracy of the classification, with nearly all modules being classified as combinational. However, after enabling the context window to be the standard length and extracting the final response in post-processing, the classification ability of the models significantly increased.
\vspace{1cm}
\begin{lstlisting}[language=Python, label = {listing:listing3}, caption={Classification prompt ($f_{\mathrm{cls2}}$) defining the optimal IR structure.},style=prettyverilog,aboveskip=10pt,firstnumber=1,linewidth=\linewidth]
You are a professional hardware engineer.
Please read the following combinational circuit description and Verilog instantiation below.
Then, determine which structure would be most helpful in determining the combinational circuits implementation.

Here are the list of possible structures:
- truth_table
- boolean_expression
- karnaugh_map
- mux_mapping
- other

Each method has its own advantages and disadvantages, and the best choice depends on the specific requirements of the design. These are listed below:
- truth_table  
  Effective in enumerating and validating all possible outputs for every input configuration. Best for very small input spaces (<= 4 inputs), since tables grow exponentially. 

- boolean_expression  
  Effective for algorithmic or pattern-based logic, especially when the input space is too large to list exhaustively. You capture the logic in concise formulas like `(a & b) | (~c & d)`.

- karnaugh_map  
  Effective for up to about 6 variables when you want minimal sum-of-products: you group adjacent 1's to derive prime implicants and get a near-optimal Boolean expression. Great when you can sketch the map and need the fewest product terms.

- mux_mapping  
  Effective when the function selects among several data inputs using one or more select signals (e.g. table-lookup, small ROMs, priority logic). You list each select pattern alongside its chosen input, then implement as a chain of multiplexers or a `case` statement.

- other
  If none of the above options fit effectively, choose this option.
  
------Beginning of High level description of Verilog------

{description}

------End of High level description of Verilog------
 
Respond with exactly one option from the list above in a code block
For example:
```truth_table```
```boolean_expression```
```karnaugh_map```
```mux_mapping```
```other```
Please do not respond with any other text in your response.
"""
\end{lstlisting}

After the first classification, the AoT process applies the secondary classification prompt in Listing~\ref{listing:listing3} in the case the module is defined as ``combinational,'' as all ``sequential'' models are given a consistent IR (FSM). The template prompt above provides brief descriptions of each combinational IR structure, and similarly prompts the LLM to respond with a specific single-term answer as its final response. The most applicable IR structure is then defined for all modules, enabling next abstraction to be applied.

\begin{lstlisting}[language=Python, label = {listing:listing4}, caption={Template prompt ($f_{\mathrm{IR}}$) — Boolean equation IR structure.},style=prettyverilog,aboveskip=10pt,firstnumber=1,linewidth=\linewidth]
You are a professional hardware engineer.
When given a Verilog module header and a brief problem description, your task is to:

1. Identify all input and output signal names from the module header.
2. Identify the relationship between the input and output signals based on the description.
3. Generate the boolean expression that describes the relationship between the input and output signals.
4. Present the result as a boolean expression, using the following format:

Respond with a JSON block only. Please do not respond with any other text in your response.
The JSON block should have the format:
```
{{
  "input": ["input_var_1", "input_var_2", ...],
  "output": ["output_var_1", "output_var_2", ...],
  "boolean_expression": "<boolean_expression>"
}}
```

------Beginning of High level description of Verilog------

{description}

------End of High level description of Verilog------

    # Example:
    Description:
    Create a Verilog module for Y = (A AND B) OR C.
    module top_module(input A, input B, input C, output Y);
    
    Response:
    ```
    {{
      "input": ["A", "B", "C"],
      "output": ["Y"],
      "boolean_expression": "(A & B) | C"
    }}
    ```
// Here are hints from previous responses. Evaluate and use them wisely:

This is a combinational circuit implementation.
This Verilog module can be solved with boolean equations.
\end{lstlisting}
The above template (Listing~\ref{listing:listing4}) is then applied in the second abstraction of AoT (generating an intermediate IR) in the case that the module is classified as a Boolean equation. The prompt defines the specific JSON structure that the LLM should respond with, followed by a 1-shot example of the intended response on a separate boolean problem. We found that providing the 1-shot example assisted the smaller models (\texttt{Llama-3.1} and \texttt{DS-Coder-V2-Lite-Instruct}) in generating higher quality IR, and is therefore implemented within each additional classification prompt. Lastly, the information gained from the prior abstraction (the classification) is appended at the end of the prompt. A similar format is utilized in the other template prompts for IR generation, listed below.

\begin{lstlisting}[language=Python, label = {listing:listing5}, caption={Template prompt ($f_{\mathrm{IR}}$) — Finite-state machine (FSM) IR structure.},style=prettyverilog,aboveskip=10pt,firstnumber=1,linewidth=\linewidth]
You are a professional hardware engineer that designs Verilog (RTL) circuit codes based on natural language descriptions and specifications.
Please read the high level description of the Verilog module below, along with the module's instantiation.

-------Beginning: High-Level Module Description--------
{description}
-------End: High-Level Module Description--------

This is a sequential circuit design, and therefore requires the need to keep track of all of the circuit's variables, or "state", at various times (clock cycles) based upon the prior states (conditions).
This can be done through a Finite-State Machine implementation.

Analyze the above description carefully, determine the following components needed to correctly implement the associated finite state machine:
- states
- transitions
- outputs

Then, generate this information for the Finite State Machine in the exact format below:

```FSM-JSON
{
  "states": ["<state1>", "<state2>", ...],
  "transitions": [
    {"from":"<state>","to":"<state>","cond":"<signal or expression>"},
    ...
  ],
  "outputs": [
    {"state":"<state>","signal":"<out_sig>","value":<0|1>},
    ...
  ]
}
```

This intermediate representation of the FSM serves as a helpful addition to the description above in defining the module.
At the end of your response, generate the final FSM result exactly in the format listed above (i.e., beginning with ```FSM-JSON and ending with ```).

# Example (one-shot)
    {example description}
    {example output}

// Here are hints from previous responses. Evaluate and use them wisely:

This is a sequential circuit implementation.
\end{lstlisting}
The template structure in Listing~\ref{listing:listing5} is applied in all cases in which the module is defined as sequential. This prompt similarly instructs the LLM to represent the circuit in a specified  JSON structure (for FSMs), defining each potential transition within the circuit. We find that this representation is successful on most small-scale circuits, however can be verbose in the event there are large numbers of states, motivating additional investigation into effective representations of sequential circuits.
\begin{lstlisting}[language=Python, label = {listing:listing6}, caption={Template prompt ($f_{\mathrm{IR}}$) — Karnaugh-Map IR structure.},style=prettyverilog,aboveskip=10pt,firstnumber=1,linewidth=\linewidth]
You are a professional hardware engineer.
When given a Verilog module header and a brief problem description, your task is to:
1. Identify all input and output signal names from the module header.
2. Enumerate every possible combination of the input signals.
3. For each combination, compute the correct output values based on the description.
4. Present the result as a JSON-formatted Karnaugh map, with input and output variables matching the signal names in the same order they appear in the module header.

Respond with a JSON block only. Please do not respond with any other text in your response.
The JSON block should have the format:
```
{{
  "input": ["input_var_1", "input_var_2", ...],
  "output": "output_var",
  "column_vars": ["input_var_1", "input_var_2", ...],
  "row_vars": ["input_var_3", "input_var_4", ...],
  "karnaugh_map": [
    [val_cell_0, val_cell_1, ...],
    [val_cell_2, val_cell_3, ...],
    ...
  ]
}}
```

------Beginning of High level description of Verilog------

{description}

------End of High level description of Verilog------

# Example:
    
    {example description}
    {example response}
    
    ```

// Here are hints from previous responses. Evaluate and use them wisely:

This is a combinational circuit implementation.
This Verilog module can be solved with karnaugh_maps.
\end{lstlisting}
The template structure in Listing~\ref{listing:listing6} is applied when combinational circuits are classified as K-map structures. In this case, the LLM is prompted with generating a condensed K-map representation of the logic, representing the output of each combination of input variables. This enables the LLM to have a structured representation of complicated logic descriptions, assisting in the translation to Verilog.
\begin{lstlisting}[language=Python, label = {listing:listing7}, caption={Template prompt ($f_{\mathrm{IR}}$) — MUX mapping IR structure.},style=prettyverilog,aboveskip=10pt,firstnumber=1,linewidth=\linewidth]

You are a professional hardware engineer.
When given a Verilog module header and a brief problem description, your task is to:
1. Identify all input and output signal names from the module header.
2. Identify the relationship between the input and output signals based on the description.
3. Generate the MUX mapping (a circuit built with multiplexers) that describes the relationship between the input and output signals.
4. Present the result as a MUX mapping, using the following JSON block format.

Respond with a JSON block only. Please do not respond with any other text in your response.
The JSON block should have the format:
```
{
  "mux_1":
  {
    "type": "mux_type_1",
    "output": "output_var_1",
    "select": ["select_var_11", "select_var_12", ...],
    "input":
    {
        value_11: "input_var_11" or value,
        value_12: "input_var_12" or value,
        ...
    }
  }
```

------Beginning of High level description of Verilog------
{description}
------End of High level description of Verilog------
    
#Example
    
    {example description}
    {example response}
    ```

// Here are hints from previous responses. Evaluate and use them wisely:

This is a combinational circuit implementation.
This Verilog module can be solved with mux_mapping.
\end{lstlisting}
In the case the LLM classifies the module as a MUX problem, Listing~\ref{listing:listing7} is applied to effectively represent the mapping between input and output signals. Given that many circuits serve to drive signals between interconnects based on specific conditions, this representation assists the LLM in defining these relationships.
\begin{lstlisting}[language=Python, label = {listing:listing8}, caption={Template prompt ($f_{\mathrm{IR}}$) — Truth table IR structure.},style=prettyverilog,aboveskip=10pt,firstnumber=1,linewidth=\linewidth]
You are a professional hardware engineer.
When given a Verilog module header and a brief problem description, your task is to:

1. Identify all input and output signal names from the module header.
2. Enumerate every possible combination of the input signals.
3. For each combination, compute the correct output values based on the description.
4. Present the result as a JSON-formatted truth table, with input and output variables matching the signal names in the same order they appear in the module header.

Respond with a JSON block only. Please do not respond with any other text in your response.
The JSON block should have the format:
```
{{
  "input": ["input_var_1", "input_var_2", ...],
  "output": ["output_var_1", "output_var_2", ...],
  "truth_table": [
    {{"input_var_1": <0|1>, "input_var_2": <0|1>, ..., "output_var_1"= <0|1>, "output_var_2"= <0|1>}},
    {{"input_var_1": <0|1>, "input_var_2": <0|1>, ..., "output_var_1"= <0|1>, "output_var_2"= <0|1>}},
    {{"input_var_1": <0|1>, "input_var_2": <0|1>, ..., "output_var_1"= <0|1>, "output_var_2"= <0|1>}},
    {{"input_var_1": <0|1>, "input_var_2": <0|1>, ..., "output_var_1"= <0|1>, "output_var_2"= <0|1>}},
    ...
  ]
}}
```

------Beginning of High level description of Verilog------

{description}

------End of High level description of Verilog------

    # Example:
    {example description}
    {example response}
    ```

    ```
    
// Here are hints from previous responses. Evaluate and use them wisely:

This is a combinational circuit implementation.
This Verilog module can be solved with truth_tables.
\end{lstlisting}
The final IR template prompt is depicted Listing~\ref{listing:listing8} is applied when the circuit is classified as a truth table structure. The LLM is prompted to enumerate through all possible combinations of inputs with their associated outputs to define the circuits intended functionality. This is optimal in circuits with small number of input combinations for an exhaustive search, however larger circuits quickly result in infeasible IR representations due to large token counts (motivating the Boolean approach in Listing~\ref{listing:listing4}).
\begin{lstlisting}[language=Python, label = {listing:listing9}, caption={Template prompt ($f_{\mathrm{ps}}$) — pseudocode abstraction.},style=prettyverilog,aboveskip=10pt,firstnumber=1,linewidth=\linewidth]

Prompt 1:  
Analyze the Verilog high-level description and instantiation below, and determine what the hardware design intends to accomplish.
Then, after careful analysis, determine the correct implementation of the Verilog module. Then in your response, generate a pseudocode representation of the Verilog module, line-by-line, which describes what each line of the Verilog module should accomplish.

------Beginning of High level description of Verilog------
{example description}
------End of High level description of Verilog------

For additional clarity, in your response, only generate succinct pseudocode comments themselves associated with each line of the verilog file.
Be sure to include specific details for key components of the design, such as the module name, and variable names/types/bit widths.
You must not generate any actual Verilog in your response.
Lastly, please do not respond with any other text or discussion - only respond with the pseudocode text itself.

Response 1:
```
{example response}
```

Prompt 2:
Analyze the Verilog high-level description and instantiation below, and determine what the hardware design intends to accomplish.
 
Then, after careful analysis, determine the correct implementation of the Verilog module. Then in your response, generate a pseudocode representation of the Verilog module, line-by-line, which describes what each line of the Verilog module should accomplish.

------Beginning of High level description of Verilog------
{description}
------End of High level description of Verilog------

For additional clarity, in your response, only generate succinct pseudocode comments themselves associated with each line 
of the verilog file.
 Be sure to include specific details for key components of the design, such as the module name, and variable names/types/bit widths.

You must not generate any actual Verilog in your response.
 Lastly, please do not respond with any other text or discussion - only respond with the pseudocode text itself.

// Here are hints from previous responses. Evaluate and use them wisely:

This is a sequential circuit implementation.

Here is the FSM representation for this Verilog module:
{JSON}

Response 2:
\end{lstlisting}
 After all circuits have been classified and generated an intermediate IR structure, the third and final abstraction of AoT is applied — line-by-line pseudocode generation. As shown in the prompt above (Listing~\ref{listing:listing9}), the LLM is prompted to generate a line-by-line pseudocode representation of the module. An example (one-shot) is included, along with all information obtained form the prior two abstractions. In this case, this includes the classification (sequential FSM) and the JSON IR representation. Upon the retrieval of the pseudocode, the final AoT prompt is constructed below.

\begin{lstlisting}[language=Python, label = {listing:listing10}, caption={Template prompt ($f_{\mathrm{final}}$)— final AoT prompt.},style=prettyverilog,aboveskip=10pt,firstnumber=1,linewidth=\linewidth]
You are a professional hardware engineer. 
Carefully analyze the description, and module declaration of the provided Verilog module.
Then, generate the code for the associated Verilog implementation of that module.

------Beginning of High level description of Verilog------
{description}
------End of High level description of Verilog------

Your final solution should be enclosed in a code block:
```verilog
<your_answer>
```
Please do not respond with any other text in your response.

// Here are hints from previous responses. Evaluate and use them wisely:

This is a combinational circuit implementation.
This Verilog module can be solved with mux_mapping.
Here is the MUX mapping for this Verilog module:

{JSON_IR}

Here is the pseudocode representation of the Verilog module:
{pseudocode}
...
\end{lstlisting}
 Finally, after the three abstractions have been applied and their information retrieved (classification, JSON IR, and pseudocode), the final AoT prompt can then be constructed as shown in Listing~\ref{listing:listing10}. Here, the LLM is prompted to generate the final Verilog representation of the model. As before, the template consists of the baseline Verilog prompt (\texttt{\{description}\}), and all prior information obtained from the abstractions. Through these representations, additional design information can be utilized within the LLM's context, thereby minimizing the complexity of the final Verilog generation.
\subsection{Alternative Inference Strategies — Implementations}
Below are additional details regarding the implementation of alternative prompting strategies utilized for comparison against AoT. This includes how we implemented the one-shot, CoT, SoT, and ToT approaches given the available (open-source) repositories, along with any limitations associated with our applications of the strategies to the Verilog generation prompts. Through this delineation, we seek to ensure fair comparisons and maintain transparency in our implementation methodologies.

\textbf{One-shot}: The one-shot strategy was implemented through prepending an example of a Verilog problem and response that contains a similar structure to the VerilogEval prompts. This example consists of a Verilog module that computes the dot-product of two 8-bit vectors, a task not included in VerilogEval dataset, to mitigate contaminating the benchmark. This same example was applied to each of the 156 VerilogEval prompts to maintain consistency in evaluation.

\textbf{CoT}: For the standard single-model evaluations, the Chain-of-Thought (CoT) inference strategy was implemented through prepending one ``exemplar'' before the baseline VerilogEval prompt. As described in the CoT work, this exemplar consists of a prompt and a series of intermediate reasoning steps that lead to the final answer, inducing the LLM to utilize a similar reasoning structure in its response. Furthermore, the problem used for the exemplar is a simple Verilog prompt not within the VerilogEval dataset (a D-latch module). Similar to the one-shot strategy, the same exemplar is prepended to all VerilogEval prompts to maintain consistency. 

However, in the multi-model evaluations, the CoT strategy is split between two models — one to generate the reasoning path, and one to generate the Verilog. Therefore, to ensure the LLM generating the CoT does not generate any true Verilog, the model is simply prompted to ``think step-by-step to determine the solution,'' rather than providing an exemplar. This implicit CoT strategy, a common alternative to explicit exemplars, enables a simple alternative for the multi-model experiments.

\textbf{SoT}: The Sketch-of-Thought (SoT) framework was implemented through applying the open-source SoT repository directly within the construction of each associated Verilog prompt. There are no alterations made to the code aside from adjusting the base prompts to the VerilogEval set.

\textbf{ToT}: The Tree-of-Thought (ToT) prompting strategy was implemented utilizing the associated open-source codebase, with modifications to support the VerilogEval benchmark. Due to the multiple inference calls required for each problem, scalability became a concern when running the benchmark multiple times. To address this, we configured ToT with a maximum tree depth of 2, two leaves per node, and one leaf retained per generation. Furthermore, after generating all responses at each depth, the LLM performs a voting step to select the best candidate. Lastly, for each generated node, we define a ``thought'' to be a full Verilog module solution to the problem, rather than splitting up the response into smaller intermediate thoughts. Although future works may explore additional optimizations of ToT for Verilog, we seek to provide a fair, base comparison against a common application of the ToT framework in terms of performance and token overhead.

\subsection{Extended Evaluations}
\textbf{Setup.} In this section, we extend our evaluations of the models and prompting strategies to include pass@5 in addition to the pass@1 for the VerilogEval-Human v1.0 benchmark, as shown in Table~\ref{tab:GroupedVerilogEval_appendix}. In these experiments, we run VerilogEval 5 times ($n=5$) for all strategies. As in the main experiments, we include the results from our best observed configuration of AoT in comparison to the alternative strategies for each model configuration. Furthermore, we extend the experiments to include a reasoning model, \texttt{GPT-o3-mini}), to examine the relative applicability of AoT to models pre-trained for reasoning tasks. This includes an additional multi-model setup, in which \texttt{GPT-o3-mini} generated abstractions are utilized with \texttt{DS-Coder-V2-Lite-Instruct} generating the final Verilog.

Moreover, we additionally expand our ablation study results to pass@5, shown in Table~\ref{tab:ablation_appendix}. Furthermore, we evaluate three additional model configurations across the AoT framework — two single model configurations (\texttt{GPT-4o} and \texttt{GPT-o3-mini}), and \texttt{o3-mini w/ DSCoder} (i.e., \texttt{DS-Coder-V2-Lite-Instruct}). Through these ablations, we seek to further identify where each component provides the most benefit to varying model types.
{\scriptsize
\setlength{\tabcolsep}{2pt}
\renewcommand{\arraystretch}{0.95}
\begin{table*}[h!]
  \caption{Evaluation of LLMs on VerilogEval-Human across Prompt Strategies (pass@1 and pass@5) (\textit{Asterisk (*) denotes single-model results for comparison due to incompatibility for multi-model inference strategies})}
  \label{tab:GroupedVerilogEval_appendix}
  \centering
  \resizebox{1.0\textwidth}{!}{%
    \begin{tabular}{cc|*{6}{c}|*{6}{c}}
      \hline
      \multicolumn{14}{c}{VerilogEval (\%)} \\
      \hline
      \multirow{2}{*}{Model} & \multirow{2}{*}{Evaluation}
        & \multicolumn{6}{c|}{pass@1 (\%)}
        & \multicolumn{6}{c}{pass@5 (\%)} \\
      \cline{3-14}
      & 
        & Baseline & 1‑shot & CoT   & SoT   & ToT   & AoT
        & Baseline & 1‑shot & CoT   & SoT   & ToT   & AoT \\
      \hline
      
        \multirow{2}{*}{GPT‑4o}
          & Compilation
            & 72.3    & 71.7  & 78.3    & 75.4  & 77.4    & \cellcolor{blue!20}\textbf{80.9}
            & 78.8           & 77.6  & 83.3    & 85.3  & 83.3    & \cellcolor{blue!20}\textbf{86.5} \\
          & Functionality
            & 57.8    & 56.2  & 59.0    & 56.3  & 60.1    & \cellcolor{blue!20}\textbf{60.4}
            & 66.7           & 64.1  & 67.9    & \cellcolor{blue!20}\textbf{69.2}  & 67.9    & 66.7 \\
          \hline
      \multirow{2}{*}{GPT‑4o‑mini}
      & Compilation
        & 67.1    & 69.5  & 69.9           & 12.6  & 62.8    & \cellcolor{blue!20}\textbf{74.9}
        & 75.0    & 78.2  & 76.9           & 28.2  & 78.2    & \cellcolor{blue!20}\textbf{82.7} \\
      & Functionality
        & \cellcolor{blue!20}\textbf{48.3}  & 48.1  & 46.2    & 6.9   & 40.8    & 47.7
        & \cellcolor{blue!20}\textbf{59.6}  & 53.2  & 53.8    & 16.0  & 56.4    & 56.4 \\
      \hline        
      \multirow{2}{*}{DS‑Coder‑V2‑Instruct}
      & Compilation
        & 79.9    & 72.2  & \cellcolor{blue!20}\textbf{80.6}  & 70.9  & 77.1    & 76.3
        & \cellcolor{blue!20}\textbf{91.0}   & 86.5  & 88.5  & 84.6  & 86.5    & 89.7 \\
      & Functionality
        & 46.9    & 43.8  & \cellcolor{blue!20}\textbf{49.7}  & 44.2  & 45.3    & 40.4
        & 58.3    & 54.5  & \cellcolor{blue!20}\textbf{59.0}  & 53.8  & 55.1    & 56.4 \\
      \hline

      \multirow{2}{*}{Llama‑3.1‑8B}
      & Compilation
        & 41.0    & 31.9  & \cellcolor{blue!20}\textbf{52.7}  & 23.8  & 39.1    & 47.9
        & 70.5    & 51.3  & \cellcolor{blue!20}\textbf{87.2}  & 61.5  & 83.9    & 84.0 \\
      & Functionality
        & 16.2    & 13.1  & \cellcolor{blue!20}\textbf{21.9}  & 7.1   & 9.62    & 13.7
        & 28.8    & 23.7  & \cellcolor{blue!20}\textbf{41.0}  & 18.6  & 29.5    & 31.4 \\
      \hline
      \multirow{2}{*}{GPT-o3‑mini}
      & Compilation
        & \cellcolor{blue!20}\textbf{86.8}  & 82.8  & 85.4    & 61.7  & 75.1    & 86.4
        & 93.6           & 91.0  & \cellcolor{blue!20}\textbf{94.2}  & 85.3  & 92.3    & 93.6 \\
      & Functionality
        & \cellcolor{blue!20}\textbf{74.6}  & 69.3  & 73.8    & 50.5  & 65.4    & 69.2
        & \cellcolor{blue!20}\textbf{84.0}  & 80.1  & 82.1    & 74.4  & 82.1    & 80.8 \\
      \hline

\multirow{2}{*}{o3-mini \& DS-Coder-V2-Lite-Instruct}
      & Compilation
        & \cellcolor{blue!20}\textbf{79.9*}    & 72.2*  & 79.4    & 70.9*  & 78.2    & \cellcolor{blue!20}\textbf{79.9}
        & \cellcolor{blue!20}\textbf{91.0*}    & 86.5*  & 87.2    & 84.6*  & 87.8      & \cellcolor{blue!20}\textbf{91.0} \\
      & Functionality
        & 46.9*    & 43.8*  & 59.7    & 44.2*  & 50.0    & \cellcolor{blue!20}\textbf{65.4}
        & 58.3*    & 54.5*  & 71.8   & 53.8*  & 57.7      & \cellcolor{blue!20}\textbf{73.7} \\
      \hline

\multirow{2}{*}{4o-mini \& DS-Coder-V2-Lite-Instruct}
      & Compilation
        & 79.9*    & 72.2*  & 79.3    & 70.9*  & 78.3    & \cellcolor{blue!20}\textbf{80.1}
        & \cellcolor{blue!20}\textbf{91.0*}    & 86.5*  & 89.1    & 84.6*  & 86.5    & \cellcolor{blue!20}\textbf{91.0} \\
      & Functionality
        & 46.9*    & 43.8*  & 50.4    & 44.2*  & 47.3    & \cellcolor{blue!20}\textbf{51.5}
        & 58.3*    & 54.5*  & \cellcolor{blue!20}\textbf{60.9}    & 53.8*  & 56.4      &  \cellcolor{blue!20}\textbf{60.9} \\
      \hline

\multirow{2}{*}{4o-mini \& Llama-3.1-8B-Instruct}
      & Compilation
        & 41.0*  & 31.9* & 52.8  & 23.8*  &  51.9   & \cellcolor{blue!20}\textbf{68.5}
        & 70.5*  &   51.3*    &   82.1    &     61.5*   &  82.1   &  \cellcolor{blue!20}\textbf{89.7}\\
      & Functionality
        & 16.2*   & 13.1* & 25.4  & 7.1*  &  14.1   &  \cellcolor{blue!20}\textbf{35.9}     
        & 28.8*   &    23.7*   &  42.3     &   18.6*    &  30.8   &  \cellcolor{blue!20}\textbf{49.4}\\
      \hline
      
    \end{tabular}
  }
\end{table*}
}

{\scriptsize
\setlength{\tabcolsep}{2pt}
\renewcommand{\arraystretch}{0.9}
\begin{table*}[h!]
  \caption{Ablation Study on AoT Components on VerilogEval-Human (\textit{Asterisk (*) denotes single-model results for comparison due to incompatibility for multi-model inference strategies})}
  \label{tab:ablation_appendix}
  \centering
  \resizebox{\textwidth}{!}{%
    \begin{tabular}{cc|*{6}{c}|*{6}{c}}
      \hline
      \multirow{2}{*}{Model} & \multirow{2}{*}{Evaluation}
        & \multicolumn{6}{c|}{pass@1 (\%)}
        & \multicolumn{6}{c}{pass@5 (\%)} \\
      \cline{3-14}
      & 
        & Base & Pseudo & Base + Pseudo & IR
        & Base + IR & Full
        & Base & Pseudo & Base + Pseudo & IR
        & Base + IR & Full \\
      \hline

      \multirow{2}{*}{GPT‑4o}
      & Comp.
        & 72.3 & 74.2 & 80.1 & 57.6 & 80.4 & \cellcolor{blue!20}\textbf{80.9}
        & 78.8 & 84.6 & \cellcolor{blue!20}\textbf{87.8} & 73.1 & 85.9 & 86.5 \\
      & Func.
        & 57.8 & 53.1 & \cellcolor{blue!20}\textbf{60.4} & 33.8 & \cellcolor{blue!20}\textbf{60.4} & 59.7
        & 66.7 & 64.7 & \cellcolor{blue!20}\textbf{69.2} & 44.2 & \cellcolor{blue!20}\textbf{69.2} & 66.7 \\
      \hline

      \multirow{2}{*}{GPT‑o3‑mini}
      & Comp.
        & 86.8 & 85.0 & 89.2 & 56.5 & \cellcolor{blue!20}\textbf{89.9 }& 86.4
        & 93.6 & 91.0 & 93.6 & 71.8 & \cellcolor{blue!20}\textbf{97.4} & 93.6 \\
      & Func.
        & \cellcolor{blue!20}\textbf{74.6} & 71.4 & 74.4 & 43.6 & 74.0 & 69.2
        &\cellcolor{blue!20}\textbf{84.0} & 81.4 & 80.8 & 57.1 & \cellcolor{blue!20}\textbf{84.0 }& 80.8 \\
      \hline

      \multirow{2}{*}{o3-mini w/ DSCoder}
      & Comp.
        & 79.9* & 75.0 & 80.5 & 49.0 & 77.6 & \cellcolor{blue!20}\textbf{82.3}
        &\cellcolor{blue!20}\textbf{91.0*} & 82.1 & 88.5 & 69.2 & 84.6 & 87.2 \\
      & Func.
        & 46.9* & 60.6 & 63.6 & 24.0 & 49.4 & \cellcolor{blue!20}\textbf{65.4}
        & 58.3* & 72.4 & \cellcolor{blue!20}\textbf{73.7} & 35.3 & 58.3 & 71.2 \\
      \hline
      
      \multirow{2}{*}{4o-mini w/ DSCoder}
      & Comp.
        & 79.9* & 69.7 & 78.1 & 57.8 & 78.3 & \cellcolor{blue!20}\textbf{80.1}
        & \cellcolor{blue!20}\textbf{91.0*} & 81.4 & 85.9 & 73.1 & 86.5 & \cellcolor{blue!20}\textbf{91.0} \\
      & Func.
        & 46.9* & 47.1 & 50.6 & 21.4 & 45.9 & \cellcolor{blue!20}\textbf{51.5}
        & 58.3* & 58.3 & \cellcolor{blue!20}\textbf{60.9} & 31.4 & 55.8 & 59.0 \\
      \hline

    \multirow{2}{*}{4o-mini w/ Llama-3.1-8B}
      & Comp.
        & 41.0* & 67.1 & 68.1 &  45.6 & 44.2  & \cellcolor{blue!20}\textbf{68.5}
        & 70.5* & 88.5  & 89.1  & 77.6 & 71.2 & \cellcolor{blue!20}\textbf{90.0} \\
      & Func.
        & 16.2* & \cellcolor{blue!20}\textbf{35.9} & 32.4 & 16.4 & 18.7 & 33.2
        & 28.8* & \cellcolor{blue!20}\textbf{55.8}  & 47.4 & 25.6 & 31.4 & 49.3\\
      \hline

    \end{tabular}%
  }
\end{table*}
}

\textbf{Results. }From the results in Table~\ref{tab:GroupedVerilogEval_appendix}, we observe the following primary trends. In the three multi-model configurations, the AoT framework continues to exceed alternative strategies at the pass@5 metric, demonstrated by the compilability and functionality exceeding (or matching) all alternative methods for each configuration. Notably, the \texttt{o3-mini w/ DS-Coder} setup demonstrates an improvement of up to 5.7\% over the best alternative multi-modal strategy (CoT), along with a 25\% improvement over the\texttt{ DS-Coder} (single-model) AoT equivalent. However, we see that when AoT is applied to the reasoning model on its own (\texttt{GPT-o3-mini}), the performance degrades from the baseline (as do all other alternative prompting strategies). Through these observations, we can conclude that multi-model strategies with AoT continue to demonstrate success over alternative methods (primarily with abstractions generated by reasoning models), while all inference strategies (including AoT) are largely ineffective when utilizing pre-trained reasoning models. 

In the ablation study, we observe that the reasoning model (\texttt{GPT-o3-mini}) on its own does not derive improvement over baseline prompting with any AoT configuration, reinforcing the observation above that SOTA reasoning models are not an optimal application of AoT. Furthermore, we see that the full AoT framework demonstrates the most relative improvement in pass@1, while intermediate AoT versions (i.e., base + pseudo, base + IR) show more relative success in the pass@5 metric. Lastly, the \texttt{o3-mini w/ DSCoder} configuration again shows success with the full framework (pass@1) by exceeding all intermediate configurations, demonstrating the importance of abstraction quality when concatenating them within the full AoT framework.
\subsection{Statistical Significance}
\textbf{Setup.} We further assess the variability of our pass@1 results by computing the sample standard deviation over \(n = 5\) independent runs. Let \(x_i\) denote the pass@1 rate, either for compilability or functionality, in run \(i\), for \(i = 1,\dots,n\). The sample mean is defined as:

\[
\bar{x} = \frac{1}{n} \sum_{i=1}^n x_i,
\]

and the sample standard deviation is:

\[
s = \sqrt{\frac{1}{n - 1}\sum_{i=1}^n \bigl(x_i - \bar{x}\bigr)^2}.
\]

Table~\ref{tab:VerilogEval_pass1_concise} reports \(\bar{x}\pm s\) for both compilation and functionality, and Table~\ref{tab:TokenUsage_concise_appendix} shows the corresponding statistics for input and output token usage. Through these statistics, we seek to assess how reliable our metrics are in differentiating performance between inference strategies.

\textbf{Results.} In evaluating the variances depicted in Table~\ref{tab:VerilogEval_pass1_concise}, we can see that across all models, the sample standard deviation of the AoT framework does not exceed 3.0\% pass@1 in compilability or functionality (ranging from 0.6\% to 3.0\%). Furthermore, we see that for 7 of the 8 model configurations, there is at least one alternative strategy that has a higher sample standard deviation (with the exception being \texttt{4o-mini w/ Llama-3.1}). This indicates that the AoT performance is relatively consistent in performance when compared to alternative methods, likely due to the extended context provided by the abstractions, minimizing the variability in potential responses. Furthermore, we see that for 2 of the 3 multi-model configurations (\texttt{o3-mini w/ DS-Coder} and \texttt{4o-mini w/ Llama-3.1}), the AoT functionality exceeds the second best strategy by well over one standard deviation. This demonstrates that the success of AoT is statistically significant in multi-model implementations. In future work, extending the experiments to a greater number of iterations would further minimize the variability, providing additional confidence in the associated performance.

We can additionally analyze the variance in tokens utilized by each model and strategy combination in Table~\ref{tab:TokenUsage_concise_appendix}, resulting in the following primary observations. First, we can observe that the input tokens have zero variance in strategies such as baseline, one-shot, CoT, and SoT. This is due to these approaches being single-shot strategies, in which the prompts remain the same across all 5 iterations. Other strategies including ToT, AoT, and multi-model CoT have multiple steps in which the input tokens depend on prior responses, causing some variation. Furthermore, we see that across all prompt strategies that the reasoning model (\texttt{GPT-o3-mini}) has the largest output token length and standard deviation as a result of the intermediate reasoning tokens. Additionally, we can see that SoT has the smallest length and variability of output tokens across all strategies in most model configurations, supporting its goal of minimizing tokens. Regarding AoT, we can see that it has a larger variability over most strategies with the exception of ToT. This can be attributed to the multiple layers of prompts associated with the AoT framework, compounding the variation in output tokens generated.

\begin{table}[h!]
  \caption{VerilogEval-Human pass@1 (\%) with per‐strategy mean $\pm$ standard deviation. \textit{(Asterisk (*) denotes single-model results for comparison due to incompatibility for multi-model inference strategies)}}
  \label{tab:VerilogEval_pass1_concise}
  \centering
  \setlength{\tabcolsep}{4pt}
  \renewcommand{\arraystretch}{1.0}
  \resizebox{\columnwidth}{!}{%
    \begin{tabular}{cc*{6}{c}}
      \hline
      \multicolumn{8}{c}{VerilogEval pass@1 (\%)} \\
      \hline
      Model & Evaluation
        & \shortstack{Baseline\\($\pm$ SD)}
        & \shortstack{1-shot\\($\pm$ SD)}
        & \shortstack{CoT\\($\pm$ SD)}
        & \shortstack{SoT\\($\pm$ SD)}
        & \shortstack{ToT\\($\pm$ SD)}
        & \shortstack{AoT\\($\pm$ SD)} \\
      \hline

      \multirow{2}{*}{GPT-4o}
        & Comp
          & 72.3 ± 0.5
          & 71.7 ± 0.8
          & 78.3 ± 1.3
          & 75.4 ± 1.6
          & 77.4 ± 0.8
          & \cellcolor{blue!20}\textbf{80.9 ± 1.1} \\
        & Func
          & 57.8 ± 1.8
          & 56.2 ± 1.7
          & 59.0 ± 1.0
          & 56.3 ± 2.9
          & 60.1 ± 1.7
          & \cellcolor{blue!20}\textbf{60.4 ± 1.7} \\
      \hline

      \multirow{2}{*}{GPT-4o-mini}
        & Comp
          & 67.1 ± 1.0
          & 69.5 ± 0.9
          & 69.9 ± 1.9
          & 12.6 ± 0.4
          & 62.8 ± 1.4
          & \cellcolor{blue!20}\textbf{74.9 ± 1.9} \\
        & Func
          & \cellcolor{blue!20}\textbf{48.3 ± 1.3}
          & 48.1 ± 0.8
          & 46.2 ± 1.0
          & 6.9  ± 1.2
          & 40.8 ± 2.5
          & 47.7 ± 1.0 \\
      \hline

      \multirow{2}{*}{DS-Coder-V2-Instruct}
        & Comp
          & 79.9 ± 2.5
          & 72.2 ± 1.2
          & \cellcolor{blue!20}\textbf{80.6 ± 1.3}
          & 70.9 ± 1.9
          & 77.1 ± 1.9
          & 76.3 ± 2.9 \\
        & Func
          & 46.9 ± 2.3
          & 43.8 ± 1.7
          & \cellcolor{blue!20}\textbf{49.7 ± 1.8}
          & 44.2 ± 1.0
          & 45.3 ± 2.1
          & 40.4 ± 2.4 \\
      \hline

      \multirow{2}{*}{Llama-3.1-8B}
        & Comp
          & 41.0 ± 2.0
          & 31.9 ± 1.9
          & \cellcolor{blue!20}\textbf{52.7 ± 3.0}
          & 23.8 ± 2.7
          & 39.1 ± 3.2
          & 47.9 ± 2.9 \\
        & Func
          & 16.2 ± 2.2
          & 13.1 ± 2.4
          & \cellcolor{blue!20}\textbf{21.9 ± 3.2}
          & 7.1  ± 1.8
          & 9.62 ± 2.4
          & 13.7 ± 1.2 \\
      \hline

      \multirow{2}{*}{o3-mini}
        & Comp
          & \cellcolor{blue!20}\textbf{86.8 ± 1.2}
          & 82.8 ± 1.2
          & 85.4 ± 2.2
          & 61.7 ± 4.4
          & 75.1 ± 1.4
          & 86.4 ± 1.6 \\
        & Func
          & \cellcolor{blue!20}\textbf{74.6 ± 2.4}
          & 69.3 ± 1.2
          & 73.8 ± 2.5
          & 50.5 ± 3.6
          & 65.4 ± 0.9
          & 69.2 ± 2.4 \\
      \hline

      \multirow{2}{*}{o3-mini w/ DS-Coder}
        & Comp
          & 79.9* ± 2.5*
          & 72.2* ± 1.2*
          & 79.4  ± 3.3
          & 70.9* ± 1.9*
          & {78.2 ± 2.0}
          & \cellcolor{blue!20}\textbf{79.9  ± 1.1} \\
        & Func
          & 46.9* ± 2.3*
          & 43.8* ± 1.7*
          & 59.7  ± 1.8
          & 44.2* ± 1.0*
          & 50.0  ± 2.1
          & \cellcolor{blue!20}\textbf{65.4 ± 0.6} \\
      \hline

      \multirow{2}{*}{4o-mini w/ DS-Coder}
        & Comp
          & 79.9* ± 2.5*
          & 72.2* ± 1.2*
          & 79.3  ± 3.1
          & 70.9* ± 1.9*
          & 78.3  ± 2.0
          & \cellcolor{blue!20}\textbf{80.1 ± 2.5} \\
        & Func
          & 46.9* ± 2.3*
          & 43.8* ± 1.7*
          & 50.4  ± 1.1
          & 44.2* ± 1.0*
          & 47.3  ± 1.7
          & \cellcolor{blue!20}\textbf{51.5 ± 2.2} \\
      \hline

      \multirow{2}{*}{4o-mini w/ Llama-3.1}
        & Comp
          & 41.0* ± 2.0*
          & 31.9* ± 1.9*
          & 52.8  ± 4.1
          & 23.8* ± 2.7*
          & 51.9  ± 2.1
          & \cellcolor{blue!20}\textbf{68.5 ± 2.7} \\
        & Func
          & 16.2* ± 2.2*
          & 13.1* ± 2.4*
          & 25.4  ± 2.9
          & 7.1*  ± 1.8*
          & 14.1  ± 0.5
          & \cellcolor{blue!20}\textbf{35.9 ± 3.0} \\
      \hline

    \end{tabular}%
  }
\end{table}

\begin{table}[h!]
  \caption{Average Token Usage of LLMs across Prompt Strategies \textit{(Asterisk (*) denotes single-model results for comparison due to incompatibility for multi-model inference strategies)}}
  \label{tab:TokenUsage_concise_appendix}
  \centering
  \resizebox{\columnwidth}{!}{%
    \begin{tabular}{cc*{7}{c}}
      \hline
      Model & Token Type
        & \shortstack{Baseline\\($\pm$ SD)}
        & \shortstack{1-shot\\($\pm$ SD)}
        & \shortstack{CoT\\($\pm$ SD)}
        & \shortstack{SoT\\($\pm$ SD)}
        & \shortstack{ToT\\($\pm$ SD)}
        & \shortstack{AoT\\($\pm$ SD)}
        & \shortstack{AoT (per abst.)\\($\pm$ SD)} \\
      \hline

      \multirow{2}{*}{GPT-4o}
        & Input
          & \cellcolor{blue!20}\textbf{180 ± 0.0}
          & 669 ± 0.0
          & 572 ± 0.0
          & 914 ± 0.0
          & 4962 ± 81.2
          & 3236 ± 21.3
          & 1079 ± 7.1 \\
        & Output
          & 501 ± 3.4
          & 376 ± 5.4
          & 491 ± 2.2
          & \cellcolor{blue!20}\textbf{200 ± 2.1}
          & 2213 ± 35.2
          & 886 ± 23.3
          & 295 ± 7.8 \\
      \hline

      \multirow{2}{*}{GPT-4o-mini}
        & Input
          & \cellcolor{blue!20}\textbf{180 ± 0.0}
          & 669 ± 0.0
          & 572 ± 0.0
          & 914 ± 0.0
          & 5080 ± 61.7
          & 3172 ± 89.5
          & 1057 ± 29.8 \\
        & Output
          & 623 ± 7.2
          & 502 ± 9.4
          & 609 ± 8.9
          & \cellcolor{blue!20}\textbf{139*} ± 5.7
          & 2671 ± 27.1
          & 1018 ± 45.6
          & 339 ± 15.2 \\
      \hline

      \multirow{2}{*}{DS-Coder}
        & Input
          & \cellcolor{blue!20}\textbf{194 ± 0.0}
          & 738 ± 0.0
          & 643 ± 0.0
          & 1055 ± 0.0
          & 5216 ± 61.4
          & 1404 ± 227.5
          & 468 ± 75.8 \\
        & Output
          & 601 ± 17.0
          & 420 ± 18.8
          & 497 ± 10.0
          & \cellcolor{blue!20}\textbf{349} ± 8.2
          & 2455 ± 36.2
          & 476 ± 78.8
          & 159 ± 26.3 \\
      \hline

      \multirow{2}{*}{Llama-3.1-8B}
        & Input
          & \cellcolor{blue!20}\textbf{209 ± 0.0}
          & 703 ± 0.0
          & 602 ± 0.0
          & 949 ± 0.0
          & 8164 ± 234.3
          & 3524 ± 46.4
          & 1175 ± 15.5 \\
        & Output
          & 535 ± 13.7
          & 503 ± 19.7
          & 546 ± 30.7
          & \cellcolor{blue!20}\textbf{482 ± 7.4}
          & 3244 ± 175.7
          & 1762 ± 59.1
          & 587 ± 19.8 \\
      \hline

      \multirow{2}{*}{o3-mini}
        & Input
          & \cellcolor{blue!20}\textbf{ 179 ± 0.0}
          & 668 ± 0.0
          & 571 ± 0.0
          & 913 ± 0.0
          & 5432 ± 96.2
          & 3362 ± 19.8
          & 1121 ± 6.6 \\
        & Output
          & 1956 ± 60.0
          & 1855 ± 33.3
          & 3997 ± 148.4
          & \cellcolor{blue!20}\textbf{1484 ± 38.1}
          & 20936 ± 394.4
          & 4576 ± 131.9
          & 1525 ± 44.0 \\
      \hline

      \multirow{2}{*}{o3-mini w/ DS-Coder}
        & Input
          & \cellcolor{blue!20}\textbf{ 194* ± 0.0*}
          & 738* ± 0.0*
          & 885 ± 8.6
          & 1055* ± 0.0*
          & 5175 ± 149.3
          & 3317 ± 0.0
          & 1106 ± 0.0 \\
        & Output
          & 601* ± 17.0*
          & 420* ± 18.8*
          & 395 ± 18.5
          & \cellcolor{blue!20}\textbf{349* ± 8.2*}
          & 9025 ± 177.0
          & 12135 ± 0.0
          & 4045 ± 0.0 \\
      \hline

      \multirow{2}{*}{4o-mini w/ DS-Coder}
        & Input
          & \cellcolor{blue!20}\textbf{194* ± 0.0*}
          & 738* ± 0.0*
          & 861 ± 3.9
          & 1055* ± 0.0*
          & 5068 ± 137.6
          & 3257 ± 40.6
          & 1086 ± 13.5 \\
        & Output
          & 601* ± 17.0*
          & 420* ± 18.8*
          & 398 ± 5.3
          & \cellcolor{blue!20}\textbf{349* ± 8.2*}
          & 2426 ± 59.3
          & 1046 ± 31.9
          & 349 ± 10.6 \\
      \hline

      \multirow{2}{*}{4o-mini w/ Llama 3.1B}
        & Input
          & \cellcolor{blue!20}\textbf{209* ± 0.0*}
          & 703* ± 0.0*
          & 835 ± 1.9
          & 949* ± 0.0*
          & 5154 ± 177.3
          & 3173 ± 21.6
          & 1058 ± 7.2 \\
        & Output
          & 535* ± 13.7*
          & 503* ± 19.7*
          & \cellcolor{blue!20}\textbf{457 ± 21.2}
          & 482* ± 7.4*
          & 2383 ± 128.5
          & 1131 ± 88.4
          & 377 ± 29.5 \\
      \hline

    \end{tabular}%
  }
\end{table} 

\vspace{1cm} 

\subsection{Impact Statement}

The Abstraction‑of‑Thought (AoT) framework harnesses multi‑level LLM prompting to lower the barrier to hardware design through leveraging abstraction. This can enable rapid prototyping from high‑level specifications to low‑level Verilog, accelerate time‑to‑market, improve the quality of LLM‑generated hardware designs, and empower smaller teams and educational settings to utilize LLMs in an accessible format for integrated circuit design. By formalizing domain‑informed abstraction stages, AoT also fosters reproducibility, knowledge transfer, and collaborative workflows across research and industry. Its modular prompt templates can be adapted to a broad range of hardware architectures, promoting innovation and and potential applications in training reasoning-based models for hardware design. Given that care must be taken to guard against LLM hallucinations in critical circuit designs and that compute costs associated with LLM inferencing should be managed, the optimized inference strategies of AoT can support these goals through more effective LLM utilization.

\end{document}

%% file: settings.tex
\usepackage{soul}
\newcommand{\drop}[1]{\textcolor{red}{#1}}
\renewcommand{\drop}[1]{}

\usepackage{multirow}

\usepackage{enumitem}

\usepackage{tabularx}
\usepackage{cite}
\usepackage{amsmath,amssymb,amsfonts}
\usepackage{graphicx}
\usepackage{textcomp}
\usepackage[table]{xcolor}
\usepackage{lipsum}
\def\BibTeX{{\rm B\kern-.05em{\sc i\kern-.025em b}\kern-.08em
    T\kern-.1667em\lower.7ex\hbox{E}\kern-.125emX}}
    
\usepackage{booktabs} 
\usepackage{pifont}
\usepackage{mathtools}

\usepackage[ruled, vlined, norelsize]{algorithm2e} 

\SetKwInput{KwInput}{Input}
\SetKwInput{KwOutput}{Output}
\SetKwInput{KwInit}{Initialization}
\SetKwInput{Kwprocedure}{Procedure}
\usepackage{lipsum}
\usepackage{dblfloatfix}
\usepackage{algpseudocode}
\usepackage{bm}

\definecolor{cadmiumgreen}{rgb}{0.0, 0.42, 0.24}

\usepackage{verbatim}
\usepackage{listings}
\lstdefinelanguage{Verilog}{
  morekeywords={module, endmodule, input, output, reg, wire, always, begin, end, if, else, for, while, case, default},
  sensitive=false,
  morecomment=[l]{//},
  morecomment=[s]{/*}{*/},
  morestring=[b]",
}

\definecolor{shadecolor}{rgb}{0.9,0.9,0.9}
    \usepackage[preprint]{neurips_2025}

\usepackage[utf8]{inputenc} 
\usepackage[T1]{fontenc}    
\usepackage{hyperref}       
\usepackage{url}            
\usepackage{booktabs}       
\usepackage{amsfonts}       
\usepackage{nicefrac}       
\usepackage{microtype}      
\usepackage{tcolorbox}
\tcbuselibrary{skins,breakable} 
\usepackage{listings}

\usepackage[utf8]{inputenc}
\usepackage[T1]{fontenc}
\usepackage{listings}
\usepackage{wrapfig}

%% file: listing_style.tex
\definecolor{codegreen}{rgb}{0,0.6,0}
\definecolor{codegray}{rgb}{0.5,0.5,0.5}
\definecolor{codepurple}{rgb}{0.58,0,0.82}
\definecolor{backcolour}{rgb}{0.95,0.95,0.92}

\let\othelstnumber=\thelstnumber
\def\createlinenumber#1#2{
    \edef\thelstnumber{%
        \unexpanded{%
            \ifnum#1=\value{lstnumber}\relax
              #2%
            \else}%
        \expandafter\unexpanded\expandafter{\thelstnumber\othelstnumber\fi}%
    }
    \ifx\othelstnumber=\relax\else
      \let\othelstnumber\relax
    \fi
}

\lstdefinestyle{customc}{
  belowcaptionskip=1\baselineskip,
  breaklines=true,
  frame=single,
  xleftmargin=0.35cm,
  xrightmargin=0.15cm,
  numbers=left,
  numbersep=5pt,  
  language=C,
  showstringspaces=false,
  basicstyle=\footnotesize\ttfamily,
  keywordstyle=\bfseries\color{green!40!black},
  commentstyle=\itshape\color{purple!40!black},
  identifierstyle=\color{blue},
  stringstyle=\color{orange},
}

\lstdefinestyle{customcArianeExploit1}{
  breaklines=true,
  frame=single,
  xleftmargin=0.4cm,
  xrightmargin=0.2cm,
  numbers=left,
  numbersep=5pt,  
  language=C,
  showstringspaces=false,
  basicstyle=\footnotesize\ttfamily,
  keywordstyle=\bfseries\color{green!40!black},
  commentstyle=\itshape\color{purple!60!black},
  identifierstyle=\color{blue},
  stringstyle=\color{yellow!50!black},
  morekeywords={asm},
  keywordstyle=[2]\bfseries\color{brown!60!black},
}

\lstdefinestyle{customcArianeExploit}{
  breaklines=true,
  frame=single,
  xleftmargin=0.4cm,
  xrightmargin=0.2cm,
  numbers=left,
  numbersep=5pt,  
  language=C,
  showstringspaces=false,
  basicstyle=\footnotesize\ttfamily,
  keywordstyle=\bfseries\color{blue},
  commentstyle=\itshape\color{green!50!black},
  identifierstyle=\color{black},
  stringstyle=\color{brown},
  morekeywords={asm},
  keywordstyle=[2]\bfseries\color{black},
}

\lstdefinestyle{customlog}{
  breaklines=true,
  frame=single,
  xleftmargin=0.35cm,
  xrightmargin=0.15cm,
  numbers=left,
  numbersep=5pt,  
  language=C,
  showstringspaces=false,
  basicstyle=\footnotesize\ttfamily,
  keywordstyle=\color{blue},
  commentstyle=\itshape\color{purple!40!black},
  identifierstyle=\color{blue},
  stringstyle=\color{orange},
  keywords=[2]{INFO},
  keywords=[3]{ERROR},x
  keywordstyle=[2]\bfseries\color{green!40!black},
  keywordstyle=[3]\bfseries\color{red!500!black},
}

\definecolor{verilogcommentcolor}{RGB}{0,124,0}
\definecolor{verilogkeywordcolor}{RGB}{49,49,255}
\definecolor{backcolor}{RGB}{237,237,237}
\definecolor{verilogsystemcolor}{RGB}{128,0,255}
\definecolor{verilognumbercolor}{RGB}{255,143,102}
\definecolor{verilogstringcolor}{RGB}{160,160,160}
\definecolor{verilogdefinecolor}{RGB}{128,64,0}
\definecolor{verilogoperatorcolor}{RGB}{0,0,128}
\definecolor{pointcolor}{RGB}{192,0,0} 
\lstdefinestyle{prettyverilog}{
   backgroundcolor=\color{backcolor},
   language           = Verilog,
   commentstyle       = \color{verilogcommentcolor},
   alsoletter         = \$'0123456789\`,
   literate           = *{+}{{\verilogColorOperator{+}}}{1}%
                         {-}{{\verilogColorOperator{-}}}{1}%
                         {@}{{\verilogColorOperator{@}}}{1}%
                         {;}{{\verilogColorOperator{;}}}{1}%
                         {*}{{\verilogColorOperator{*}}}{1}%
                         {?}{{\verilogColorOperator{? }}}{1}%
                         {:}{{\verilogColorOperator{:}}}{1}%
                         {<}{{\verilogColorOperator{<}}}{1}%
                         {>}{{\verilogColorOperator{>}}}{1}%
                         {!}{{\verilogColorOperator{!}}}{1}%
                         {xorsymbol}{{\verilogColorOperator{^}}}{1}%
                         {|}{{\verilogColorOperator{| }}}{1}%
                         {||}{{\verilogColorOperator{|| }}}{1}%
                         {=}{{\verilogColorOperator{= }}}{1}%
                         {==}{{\verilogColorOperator{== }}}{1}%
                         {=>}{{\verilogColorOperator{=> }}}{1}%
                         {[}{{\verilogColorOperator{[}}}{1}%
                         {]}{{\verilogColorOperator{]}}}{1}%
                         {(}{{\verilogColorOperator{(}}}{1}%
                         {)}{{\verilogColorOperator{)}}}{1}%
                         {rightbracket}{{\verilogColorOperator{)}}}{1}%
                         {,}{{\verilogColorOperator{,}}}{1}%
                         {.}{{\verilogColorOperator{.}}}{1}%
                         {~}{{\verilogColorOperator{$\sim$}}}{1}%
                         {\%}{{\verilogColorOperator{\%}}}{1}%
                         {\&}{{\verilogColorOperator{\& }}}{1}%
                         {\&\&}{{\verilogColorOperator{\&\& }}}{1}%
                         {\#}{{\verilogColorOperator{\#}}}{1}%
                         {\ /\ }{{\verilogColorOperator{\ /\ }}}{3}%
                         {\ _}{\ \_}{2}%
                        ,
   morestring         = [s][\color{verilogstringcolor}]{"}{"},%
   identifierstyle    = \color{black},
   vlogdefinestyle    = \color{verilogdefinecolor},
   vlogconstantstyle  = \color{verilognumbercolor},
   vlogsystemstyle    = \color{verilogsystemcolor},
   basicstyle         = \small\fontencoding{T1}\ttfamily,
  columns=fullflexible, 
   keywordstyle       = \bfseries\color{verilogkeywordcolor},
   morekeywords      = {val, when, port, coverage, unique},
   numbers            = left,
   numbersep          = 5pt,
   tabsize            = 2,
   escapeinside       = {/*!}{!*/},
   upquote            = true,
   sensitive          = true,
   showstringspaces   = false, 
   frame              = single, 
   breaklines         = true,
   abovecaptionskip   = 0pt,
   belowcaptionskip   = 0pt, 
   xleftmargin        =0.35cm,
   xrightmargin       =0.15cm,
   captionpos         = b,
   emph               = {Point, Point0, Point1, Point2, Point3, Point4, Point5, Point6, Point7, Point8, Point9},
   emphstyle          =\color{pointcolor},
   emph               = {[2] STVEC,SCOUNTEREN,MSTATUS,MTVEC,ML1_ICACHE_MISS,ML1_DCACHE_MISS,MITLB_MISS,MDTLB_MISS,
                             MLOAD,MSTORE,MEXCEPTION,MEXCEPTION_RET,MBRANCH_JUMP,MCALL,MRET,MMIS_PREDICT,MSB_FULL,
                             MIF_EMPTY,MHPM_COUNTER_17,MHPM_COUNTER_18,MHPM_COUNTER_19,MHPM_COUNTER_20,MHPM_COUNTER_21,
                             MHPM_COUNTER_22,MHPM_COUNTER_23,MHPM_COUNTER_24,MHPM_COUNTER_25,MHPM_COUNTER_26,MHPM_COUNTER_27,
                             MHPM_COUNTER_28,MHPM_COUNTER_29,MHPM_COUNTER_30,MHPM_COUNTER_31}, 
   emphstyle          = {[2]\bfseries\color{verilogkeywordcolor}}
}

\makeatletter

\newcommand\language@verilog{Verilog}
\expandafter\lst@NormedDef\expandafter\languageNormedDefd@verilog%
  \expandafter{\language@verilog}
  
\lst@SaveOutputDef{`'}\quotesngl@verilog
\lst@SaveOutputDef{``}\backtick@verilog
\lst@SaveOutputDef{`\$}\dollar@verilog

\newcommand\getfirstchar@verilog{}
\newcommand\getfirstchar@@verilog{}
\newcommand\firstchar@verilog{}
\def\getfirstchar@verilog#1{\getfirstchar@@verilog#1\relax}
\def\getfirstchar@@verilog#1#2\relax{\def\firstchar@verilog{#1}}

\newcommand\addedToOutput@verilog{}
\lst@AddToHook{Output}{\addedToOutput@verilog}

\newcommand\constantstyle@verilog{}
\lst@Key{vlogconstantstyle}\relax%
   {\def\constantstyle@verilog{#1}}

\newcommand\definestyle@verilog{}
\lst@Key{vlogdefinestyle}\relax%
   {\def\definestyle@verilog{#1}}

\newcommand\systemstyle@verilog{}
\lst@Key{vlogsystemstyle}\relax%
   {\def\systemstyle@verilog{#1}}

\newcount\currentchar@verilog
  
\newcommand\@ddedToOutput@verilog
{%
   \ifnum\lst@mode=\lst@Pmode%
      \expandafter\getfirstchar@verilog\expandafter{\the\lst@token}%
      \expandafter\ifx\firstchar@verilog\backtick@verilog
         \let\lst@thestyle\definestyle@verilog%
      \else
         \expandafter\ifx\firstchar@verilog\dollar@verilog
            \let\lst@thestyle\systemstyle@verilog%
         \else
            \expandafter\ifx\firstchar@verilog\quotesngl@verilog
               \let\lst@thestyle\constantstyle@verilog%
            \else
               \currentchar@verilog=48
               \loop
                  \expandafter\ifnum%
                  \expandafter`\firstchar@verilog=\currentchar@verilog%
                     \let\lst@thestyle\constantstyle@verilog%
                     \let\iterate\relax%
                  \fi
                  \advance\currentchar@verilog by \@ne%
                  \unless\ifnum\currentchar@verilog>57%
               \repeat%
            \fi
         \fi
      \fi
   \fi
}

\lst@AddToHook{PreInit}{%
  \ifx\lst@language\languageNormedDefd@verilog%
    \let\addedToOutput@verilog\@ddedToOutput@verilog%
  \fi
}

\newcommand{\verilogColorOperator}[1]
{%
  \ifnum\lst@mode=\lst@Pmode\relax%
   {\bfseries\textcolor{verilogoperatorcolor}{#1}}%
  \else
    #1%
  \fi
}

\makeatother

\lstdefinestyle{mystyle}{
    commentstyle=\textit,
    keywordstyle=\textbf,
    stringstyle=\color{codepurple},
    basicstyle=\ttfamily,
    breakatwhitespace=false,         
    breaklines=true,      
    frame=single, 
    framexleftmargin=\parindent,
    captionpos=b,                    
    keepspaces=true,                 
    numbers=left,    
    numberstyle=\normalsize,
    stepnumber=1,
    numbersep=5pt,   
    xleftmargin=1.5\parindent,
    showspaces=false,                
    showstringspaces=false,
    showtabs=false,                  
    tabsize=2
}

\lstdefinestyle{pythonstyle}{
 columns=fixed,
 numbers=left,                                        
 numberstyle=\tiny\color{gray},                       
 frame=trbl,                                        
 breaklines=true,                                     
 keywordstyle=\color[RGB]{40,40,255},                 
 numberstyle=\footnotesize\color{darkgray},
 commentstyle=\it\color[RGB]{0,96,96},                
 stringstyle=\rmfamily\slshape\color[RGB]{128,0,0},   
 showstringspaces=false,                              
 language=python,                                        
 basicstyle=\linespread{1.0}\fontsize{10bp}{10bp}\selectfont\ttfamily,                      
 inputencoding=utf8,
 extendedchars=false  
}

\lstdefinestyle{SnippetStyle}{
  basicstyle=\small\ttfamily,
  breaklines=true,
  columns=fullflexible,
  frame=none,
  upquote=true
}

\tcbset{
  snippetbox/.style={
    enhanced,
    sharp corners=all,
    arc=2mm,
    boxrule=0.5pt,
    colback=blue!5!white,
    colframe=blue!50!black,
    left=1mm, right=1mm, top=1mm, bottom=1mm,
    width=0.48\columnwidth,     
    halign=center,
  }
}

\tcbset{
  snippetboxfull/.style={
    enhanced,
    sharp corners=all,
    arc=2mm,
    boxrule=0.5pt,
    colback=blue!5!white,
    colframe=blue!50!black,
    left=1mm, right=1mm, top=1mm, bottom=1mm,
    width=1.0\columnwidth,     
    halign=center,
  }
}